\documentclass[twocolumn]{aastex62}

\hypersetup{breaklinks=true}

\usepackage{mathtools}

\received{2018 June 28}
\revised{2018 October 22}
\accepted{2018 October 22}
\submitjournal{ApJ}

\shorttitle{Binary planet formation}
\shortauthors{Chrenko et al.}

\begin{document}


\title{Binary planet formation by gas-assisted encounters of planetary embryos}

\correspondingauthor{Ond\v{r}ej Chrenko}
\email{chrenko@sirrah.troja.mff.cuni.cz}

\author[0000-0001-7215-5026]{Ond\v{r}ej Chrenko}
\affil{Institute of Astronomy, Charles University in Prague \\
  V Hole\v{s}ovi\v{c}k\'{a}ch 2, \\
  18000 Prague 8, Czech Republic}

\author[0000-0003-2763-1411]{Miroslav Bro\v{z}}
\affil{Institute of Astronomy, Charles University in Prague \\
  V Hole\v{s}ovi\v{c}k\'{a}ch 2, \\
  18000 Prague 8, Czech Republic}

\author[0000-0002-4547-4301]{David Nesvorn\'{y}}
\affil{Southwest Research Institute, Department of Space Studies \\
  1050 Walnut St., Suite 300, \\
  Boulder, CO, 80302, USA}



\begin{abstract}

We present radiation hydrodynamic simulations in which
binary planets form by close encounters in
a system of several super-Earth embryos.
The embryos are embedded in a protoplanetary disk consisting
of gas and pebbles and evolve in a
region where the disk structure supports convergent migration due to Type I torques.
As the embryos accrete pebbles, they become heated and thus affected
by the thermal torque \citep{Benitez-Llambay_etal_2015Natur.520...63B}
and the hot-trail effect \citep{Chrenko_etal_2017A&A...606A.114C} which excites
orbital eccentricities.
Motivated by findings of \cite{Eklund_Masset_2017MNRAS.469..206E}, we assume
the hot-trail effect operates also vertically and reduces the efficiency
of inclination damping. 
Non-zero inclinations allow the embryos to become closely packed
and also vertically stirred within the convergence zone.
Subsequently, close encounters of two embryos assisted by the disk gravity
  can form transient binary planets which quickly dissolve.
  Binary planets with a longer lifetime $\sim$$10^{4}\,\mathrm{yr}$
  form in 3-body interactions of a transient pair with one of the remaining embryos.
  The separation of binary components generally decreases in subsequent
encounters and due to pebble accretion until the binary merges, forming a giant planet core.
We provide an order-of-magnitude estimate of the expected occurrence rate of binary planets,
yielding one binary planet per $\simeq$$2$--$5\times10^{4}$ planetary systems.
Therefore, although rare, the binary planets may exist in exoplanetary
systems and they should be systematically searched for.

\end{abstract}



\section{Introduction}
\label{sec:intro}

Several classes of celestial objects are known to
exist in binary configurations (e.g. minor solar-system bodies,
dwarf planets, stars, etc.) -- as two bodies orbiting
their barycenter which is located exterior to their physical radii.
The existence of binaries is an important observational constraint
because a successful population synthesis model
for a given class of objects must be able to explain how binaries form,
how frequent they are, how they dynamically evolve and affect
their neighborhood.

Concerning binary objects within the scope
of planetary sciences, the richest sample is
in the population of minor solar-system bodies.
Examples can be found among the
near-Earth objects \citep[NEOs; e.g.][]{Margot_etal_2002Sci...296.1445M,Pravec_etal_2006Icar..181...63P,Scheeres_etal_2006Sci...314.1280S},
main-belt asteroids \citep[MBAs; e.g.][]{Marchis_etal_2008Icar..196...97M,Pravec_etal_2012Icar..218..125P}, Jovian Trojans
\citep[e.g.][]{Marchis_etal_2006Natur.439..565M,Sonnett_etal_2015ApJ...799..191S}
and surprisingly frequently among the Kuiper-belt objects \citep[KBOs; e.g.][]{Veillet_etal_2002Natur.416..711V,Brown_etal_2006ApJ...639L..43B,Richardson_Walsh_2006AREPS..34...47R,Noll_etal_2008Icar..194..758N}.
Formation of binary minor bodies
took place during various epochs of the Solar System.
Some binary asteroids
originate in recent breakup events
\citep{Walsh_etal_2008Natur.454..188W}, whereas the binary
KBOs were probably established early, during the planetesimal formation
\citep{Goldreich_etal_2002Natur.420..643G,Nesvorny_etal_2010AJ....140..785N,Fraser_etal_2017NatAs...1E..88F}
more than four billion years ago.

For large bodies, the
number of binary configurations suddenly drops almost to zero.
As for the known and confirmed objects, only Pluto and Charon
can be considered a binary
\citep{Christy_Harrington_1978AJ.....83.1005C,Walker_1980MNRAS.192P..47W,Lee_Peale_2006Icar..184..573L,Brozovic_etal_2015Icar..246..317B},
likely of an impact origin 
\citep{Canup_2011AJ....141...35C,McKinnon_etal_2017Icar..287....2M}.
Since Pluto and Charon were classified as {\em dwarf} planets,
the conclusion stands that planets in binary configurations
have not yet been discovered.

Given that more than 3700 exoplanets
have been confirmed up to
date\footnote{As of August 2018, according to the NASA Exoplanet Archive: \url{https://exoplanetarchive.ipac.caltech.edu/}.},
the paucity of binary planets is a
well-established characteristic of the dataset.
However, its implications for our understanding of
planet formation are unclear and maybe even
underrated at present.
Are the binary planets scarce and 
are we only unable to discover them with current methods?
Or is their non-existence a universal feature shared
by all planetary systems throughout the Galaxy?

To start addressing these
questions,
this paper discusses formation
of binary planets by 2- and 3-body encounters
of planetary embryos in protoplanetary disks,
during the phase when the gas is still abundant
and the embryos still grow by pebble accretion
\citep{Ormel_Klahr_2010A&A...520A..43O,Lambrechts_Johansen_2012A&A...544A..32L}.
We advocate that suitable conditions 
to form binary planets are achieved when orbital eccentricities
and inclinations or embryos are excited by thermal torques
related to accretion heating \citep{Chrenko_etal_2017A&A...606A.114C,Eklund_Masset_2017MNRAS.469..206E}.
Our model utilizes radiation hydrodynamics (RHD)
to account for these effects.

Although the results of this paper are preliminary in many aspects,
they demonstrate that
binary planets can exist
and it may be
only a matter of time (or method advancements) before an object like
this is discovered in one of the exoplanet search campaigns.
To motivate future observations and data mining,
we emphasize that promising methods for detections of exomoons
have been developed and applied in recent years. These include, for example,
the transit timing variations \citep[TTVs;][]{Simon_etal_2007A&A...470..727S}
and transit duration variations \citep[TDVs;][]{Kipping_2009MNRAS.392..181K},
their Bayesian analysis in the framework of direct star-planet-moon
modeling and fitting \citep{Kipping_etal_2012ApJ...750..115K},
photometric analysis of phase-folded light curves using the
scatter peak (SP) method \citep{Simon_etal_2012MNRAS.419..164S}
or the orbital sampling effect \citep[OSE;][]{Heller_2014ApJ...787...14H,Hippke_2015ApJ...806...51H},
microlensing events \citep{Han_2008ApJ...684..684H,Liebig_Wambsganss_2010A&A...520A..68L,Bennett_etal_2014ApJ...785..155B},
or asymmetric light curves due to plasma tori of hypothetic volcanic moons \citep{Ben-Jaffel_Ballester_2014ApJ...785L..30B}.

Indeed, Kepler-1625 b-i has been recently identified
as an exomoon candidate \citep{Heller_2018A&A...610A..39H,Teachey_2018AJ....155...36T}
and is waiting for a conclusive confirmation.
Moreover, \cite{Lewis_etal_2015ApJ...805...27L}
discusses that
the {\em CoRoT} target SRc01 E2 1066 can be explained as a binary gas-giant
planet, although the signal can also correspond
to a single planet transiting a star spot \citep{Erikson_etal_2012A&A...539A..14E}.
Therefore, methods similar to those listed above could be applicable when
searching for binary planets.

Our paper is organized as follows. In Section~\ref{sec:model},
we outline our RHD model. Section~\ref{sec:simulation}
describes our nominal simulation with the binary planet formation.
Planetary encounters are analyzed, as well as the influence of the gas disk.
Subsequently, we test the stability
of binary planets in several simplified models (without neighboring
embryos; without the disk; etc.). We also study
binary planet formation in a set of four additional simulations
to verify the relevance of the process.
In Section~\ref{sec:discussion},
we estimate the expected occurrence rate of binary
planets in the exoplanetary population and we also
discuss a possible role of binary planets
in planetary sciences.
Section~\ref{sec:conclusions} is devoted to conclusions.

\subsection{Definitions}
To avoid confusion, let us list several definitions which we
use throughout the rest of the paper.
\begin{itemize}
  \item {\em Binary} is a shortcut for a binary planet,
    not to be mistaken with binary stars etc.
  \item {\em Transient} (also {\em transient binary} or {\em transient pair})
    is a binary which forms by 2-body encounters
    of planetary embryos \citep[e.g.][]{Astakhov_etal_2005MNRAS.360..401A},
    in our case with the assistance of the disk gravity as we shall
    demonstrate later. We choose
      the name {\em transient} because we find the typical lifetime of these
    binaries to be of the order of one stellarcentric orbital period.
  \item {\em Hardening} \citep[e.g.][]{Hills_1975AJ.....80..809H}
    is a process during
    which the orbital energy of a binary configuration
    is dissipated and the separation of binary components decreases.
  \item {\em Stability} of a binary planet is considered if it can
      survive at least more than one stellarcentric orbital period,
      which is usual after hardening. In principle, such a binary can 
      be observed. We characterize the {\em stability} by means of the lifetime
      on which the binary components remain gravitationally bound.
  \item {\em Encounter} refers to a close encounter of 
    two and more planetary embryos (single or binary),
    when they enter one another's Hill sphere.
  \item {\em Merger} refers to a physical collision
    of two embryos. In our approximation, we replace
    the colliding embryos by a single object, assuming
    perfect merger (mass and momentum conservation).
  \item We denote orbital elements in the stellarcentric
    frame with a subscript `s' to distinguish them from
    the orbital elements of one binary component
    with respect to another (e.g. $a_{\mathrm{s}}$ is
    the stellarcentric semimajor axis but $a$ is the semimajor
    axis of the binary configuration).
\end{itemize}

\section{Radiation hydrodynamic model}
\label{sec:model}

\subsection{General overview}
The individual constituents of our model
are as follows.
First, we consider radiation transfer which is essential to properly
reproduce the disk structure \citep{Bitsch_etal_2013A&A...549A.124B} 
and to account for all components
of the Type I torque acting on low-mass planets \citep[e.g.][]{Baruteau_Masset_2008ApJ...672.1054B,
Kley_Crida_2008,Kley_etal_2009A&A...506..971K,
Lega_etal_2014MNRAS.440..683L}.

Second, we use a two-fluid approximation to include a disk of pebbles
which serves as a material reservoir 
for the accreting embryos
\citep{Ormel_Klahr_2010A&A...520A..43O,Lambrechts_Johansen_2012A&A...544A..32L,Morbidelli_Nesvorny_2012A&A...546A..18M}.

Third, we also take into account that pebbles 
heat the accreting embryos which
in turn heat the gas in their vicinity.
The migration is then modified due to the thermal torque \citep{Benitez-Llambay_etal_2015Natur.520...63B,Masset_2017MNRAS.472.4204M}
and its dynamical component -- the hot-trail effect \citep{Chrenko_etal_2017A&A...606A.114C,Eklund_Masset_2017MNRAS.469..206E,Masset_VelascoRomero_2017MNRAS.465.3175M} --
which perturbs the embryos
in a way that their orbital eccentricities are excited.
This is due to the epicyclic motion of the embryo,
which causes variations in the azimuthally
uneven distribution of the heated (and thus underdense)
gas.

The numerical modeling is done with the 
{\sc fargo\_thorin}
code\footnote{The
code is available at \url{http://sirrah.troja.mff.cuni.cz/~chrenko}.}
which was introduced and described in detail in \cite{Chrenko_etal_2017A&A...606A.114C}.
The code is based on {\sc fargo} \citep{Masset_2000A&AS..141..165M}.
The model is 2D (vertically averaged) but planets are evolved in 3D.
A number of important vertical phenomena were implemented, although
some with unavoidable approximations.

A new phenomenon implemented in this study
is the vertical hot-trail effect
described by \cite{Eklund_Masset_2017MNRAS.469..206E}
which can excite orbital inclinations. 
The inclination excitation should not occur for an isolated and non-inclined
orbit because it is quenched by the eccentricity growth
which is faster \citep{Eklund_Masset_2017MNRAS.469..206E},
however, the vertical hot-trail effect operates when a non-negligible
inclination is initially excited by some other
mechanism. For example, it can become important in a system 
of multiple embryos where close encounters
temporarily pump up the inclinations. The vertical hot trail
then starts to counteract the usual inclination damping by bending waves \citep{Tanaka_Ward_2004ApJ...602..388T}.

\subsection{Governing equations}

Gas is treated as a viscous Eulerian fluid described by
the surface density $\Sigma$, flow velocity $\vec{v}$ on the polar staggered mesh
and the internal energy $\epsilon$.
Pebble disk is represented by an inviscid and pressureless fluid
with its own surface density $\Sigma_{\mathrm{p}}$ and velocity $\vec{u}$.
We assume two-way coupling between both fluids by linear
drag terms, with the Stokes number $\tau$ calculated for the Epstein regime.

The RHD partial differential equations read
\begin{equation}
  \frac{ \partial \Sigma}{\partial t} + \nabla\cdot\left( \Sigma\vec{v} \right) = 0 \, ,
  \label{eq:continuity}
\end{equation}
\begin{equation}
  \begin{split}
  \frac{ \partial \vec{v}}{\partial t} + \vec{v}\cdot\nabla\vec{v} = 
  & - \frac{1}{\Sigma}\nabla P  + \frac{1}{\Sigma}\nabla\cdot\tensor{T} - \frac{\int{\rho\nabla\phi\,\mathrm{d}z}}{\Sigma} \\
  & + \frac{\Sigma_{\mathrm{p}}}{\Sigma}\frac{\Omega_{\mathrm{K}}}{\tau}\left(\vec{u}-\vec{v}\right) \, ,
  \end{split}
  \label{eq:navierestokes}
\end{equation}
\begin{equation}
  \begin{split}
  \frac{ \partial \epsilon}{ \partial t} + \nabla\cdot\left( \epsilon \vec{v} \right) = 
  & - P\nabla\cdot\vec{v} + Q_{\mathrm{visc}} +  Q_{\mathrm{acc}} \\
  & + \frac{2\sigma_{\mathrm{B}}}{\tau_{\mathrm{eff}}}\left(T_{\mathrm{irr}}^{4}-T^{4}\right) - 2H\nabla\cdot\vec{F} \, ,
  \end{split}
\label{eq:energy}
\end{equation}
\begin{equation}
  \frac{ \partial \Sigma_{\mathrm{p}}}{\partial t} + \nabla\cdot\left( \Sigma_{\mathrm{p}}\vec{u} \right) = -\left(\frac{\partial \Sigma_{\mathrm{p}}}{\partial t}\right)_{\mathrm{acc}} \, ,
  \label{eq:continuity_peb}
\end{equation}
\begin{equation}
  \frac{ \partial \vec{u}}{\partial t} + \vec{u}\cdot\nabla\vec{u} = - \frac{\int{\rho_{\mathrm{p}}\nabla\phi\,\mathrm{d}z}}{\Sigma_{\mathrm{p}}} - \frac{\Omega_{\mathrm{K}}}{\tau}\left(\vec{u}-\vec{v}\right) \, ,
  \label{eq:navierestokes_peb}
\end{equation}
where the individual quantities are the pressure $P$,
viscous stress tensor $\tensor{T}$ \citep[e.g][]{Masset_2002A&A...387..605M},
gas volume density $\rho$, volume density of pebbles $\rho_{\mathrm{p}}$,
coordinate $z$ perpendicular to the midplane,
gravitational potential $\phi$ of the primary and the planets, Keplerian
angular frequency $\Omega_{\mathrm{K}}$, viscous heating term $Q_{\mathrm{visc}}$
\citep{Mihalas_WeibelMihalas_1984frh..book.....M},
accretion heating term $Q_{\mathrm{acc}}$ related to the accretion sink term $(\partial \Sigma_{\mathrm{p}}/\partial t)_{\mathrm{acc}}$,
Stefan-Boltzmann constant $\sigma_{\mathrm{B}}$, effective vertical optical depth $\tau_{\mathrm{eff}}$
\citep{Hubeny_1990ApJ...351..632H}, irradiation temperature $T_{\mathrm{irr}}$
\citep{Chiang_Goldreich_1997ApJ...490..368C,Menou_Goodman_2004ApJ...606..520M,Baillie_Charnoz_2014ApJ...786...35B},
midplane gas temperature $T$ (the $\propto T^{4}$ term describes vertical cooling),
vertical pressure scale height $H$ and radiative flux $\vec{F}$.

The ideal gas state equation is used as the thermodynamic
closing relation
\begin{equation}
  P = (\gamma-1)\epsilon = \Sigma\frac{R}{\mu}T \, ,
  \label{eq:eos}
\end{equation}
where $\gamma$ is the adiabatic index, $R$ is the universal
gas constant and $\mu$ is the mean molecular weight.
Equation~(\ref{eq:eos}) has been
widely used in numerical models to 
relax inferior isothermal appproximations and to account for
the disk thermodynamics (through the energy equation)
which is important for accurate migration rates
\citep[e.g.][]{Kley_Crida_2008,Kley_etal_2009A&A...506..971K,Lega_etal_2014MNRAS.440..683L}.
We also point out that the given state equation neglects the radiation pressure
and phase transitions which is a valid assumption in low-temperature disks
\citep{DAngelo_etal_2003ApJ...599..548D}. 

Finally, the flux-limited diffusion and one-temperature
approximation are utilized to describe the in-plane
radiation transport, leading to
\begin{equation}
  \vec{F} = -D\nabla T = - \lambda_{\mathrm{lim}}\frac{16\sigma_{\mathrm{B}}T^{3}}{\rho_{0}\kappa}\nabla T \, ,
  \label{eq:fld}
\end{equation}
where $D$ denotes the diffusion coefficient, $\lambda_{\mathrm{lim}}$ is
the flux limiter according to \cite{Kley_1989A&A...208...98K},
$\rho_{0}$ is the midplane volume density and $\kappa(\rho,T)$
is the material opacity. We use the opacity by \cite{Bell_Lin_1994ApJ...427..987B} for both
the Rosseland and Planck opacities.

For completeness, we provide the accretion heating
formula
\begin{equation}
  Q_{\mathrm{acc}} = \sum\limits_{i}\frac{G{M_{\mathrm{em},i}\dot{M}_{\mathrm{em},i}}}{R_{i}S_{\mathrm{cell}}} \, ,
  \label{eq:q_acc}
\end{equation}
where the sum goes over all embryos with indices $i$, masses ${M_{\mathrm{em},i}}$,
self-consistently calculated pebble accretion rates ${\dot{M}_{\mathrm{em},i}}$ and
physical radii $R_{i}$. ${G}$ is the gravitational constant
and ${S_{\mathrm{cell}}}$ is the surface area of the cell which contains
the respective embryo and in which the heat is liberated (${Q_{\mathrm{acc}}}$ is zero
in other cells).

\subsection{Evolution of planets and inclination damping}
\label{sec:incl_damp}

Planets are evolved
on 3D orbits using the
{\sc ias15} integrator \citep{Rein_Liu_2012A&A...537A.128R,Rein_Spiegel_2015MNRAS.446.1424R}.
Planetary collisions are treated as perfect mergers.
The planet-disk interactions are calculated by
means of the vertical averaging procedure of
\cite{Muller_etal_2012A&A...541A.123M}.
The planetary potential
is adopted from \cite{Klahr_Kley_2006A&A...445..747K}, having
the smoothing length
$r_{\mathrm{sm}}=0.5R_{\mathrm{H}}$, where $R_{\mathrm{H}}$ is
the planet's Hill sphere.

When computing the torque acting on an embryo,
we do not exclude any part of the Hill sphere
because we focus on low-mass embryos \citep{Lega_etal_2014MNRAS.440..683L}.
Such an exclusion is required only when embryos exceed masses $\simeq$$10\,M_{\earth}$
and form a circumplanetary disk. This disk
should not contribute to the gas-driven torque because it comoves with the embryo
\citep{Crida_etal_2008A&A...483..325C}.
Our model ignores the torques from pebbles \citep{Benitez-Llambay_Pessah_2018ApJ...855L..28B} 
because we assume relatively low pebble-to-gas
mass ratios (less than $0.001$).
But we point out that during accretion of pebbles, we account
for the transfer of their mass and linear momentum onto the embryo.

An important ingredient when investigating 3D planetary orbits
is the inclination damping \citep[e.g.][]{Cresswell_etal_2007A&A...473..329C}.
We include the damping by using
the formula from \cite{Tanaka_Ward_2004ApJ...602..388T}.
In our case, the damping acceleration perpendicular to the disk plane
reads
\begin{equation}
  {a}_{z} =
  \begin{dcases*}
    \beta\frac{\Sigma\Omega_{\mathrm{K}}}{c_{\mathrm{s}}^{4}}\left(2A_{z}^{c}v_{z}+A_{z}^{s}z\Omega_{\mathrm{K}}\right) \, , & $I>I_{0}$ \\
    0 \, , & $I\leq I_{0}$
  \end{dcases*}
  \label{eq:damping}
\end{equation}
where $\beta=0.3$ \citep[e.g.][]{Pierens_etal_2013A&A...558A.105P}, 
$A_{z}^{c}=-1.088$ and $A_{z}^{s}=-0.871$ are fixed coefficients,
$c_{\mathrm{s}}$ is the sound speed,
$v_{z}$ is the embryo's vertical velocity and $z$ is its vertical separation
from the midplane. $\Sigma$ and $\Omega_{\mathrm{K}}$
are evaluated along the embryo's orbit.

In writing Equation (\ref{eq:damping}), we introduce a simple modification
of the Tanaka \& Ward's formula. We assume the inclination
damping does not operate when the orbital inclination $I$
is below a certain critical value $I_{0}$.
The motivation for this modification stems from the
findings of \cite{Eklund_Masset_2017MNRAS.469..206E} who
investigated the orbital evolution of a hot (accreting) planet
in a 3D radiative disk. Not only did they find the eccentricity
excitation due to the hot-trail effect, but they also described
the effect has a vertical component which can excite the inclinations.

In our 2D model with the accretion heating, the eccentricity excitation
is reproduced naturally. Regarding the inclinations,
we simply assume that the hot-trail effect operates vertically as well
and balances the inclination damping up to the value $I_{0}$.
For larger inclinations, the damping takes over and the standard
formula for ${a}_{z}$ applies.
We consider $I_{0}$ to be a free parameter of the model and
choose $I_{0}=10^{-3}\,\mathrm{rad}\simeq0.057^{\circ}$ which {is
at the lower end of} the results of \cite{Eklund_Masset_2017MNRAS.469..206E}.

\section{Simulations}
\label{sec:simulation}

\subsection{Disk model}
\label{sec:diskmodel}

\begin{figure}[]
  \centering
  \includegraphics[width=8.8cm]{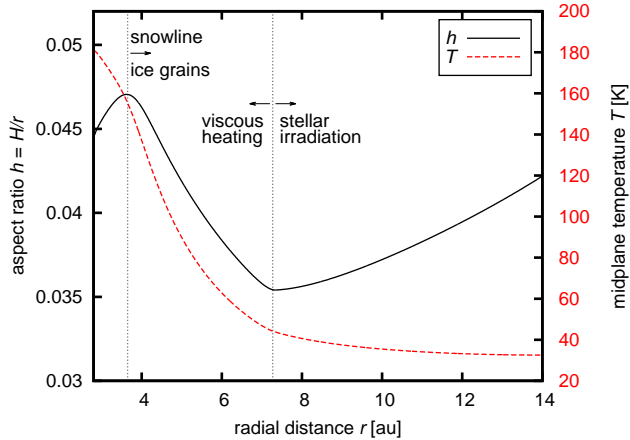}
  \caption{Radial profile of the aspect ratio $h(r)=H/r$ and temperature $T(r)$
    in the disk used in our simulations. The right vertical dotted line marks the transition
  between the viscously heated and stellar irradiated regions, the latter
  exhibiting flaring of the disk. The left vertical dotted line corresponds
  to the water ice line which is also a local maximum of the \cite{Bell_Lin_1994ApJ...427..987B}
  opacities. The figure is taken from \cite{Chrenko_etal_2017A&A...606A.114C}.}
  \label{fig:aspect}
\end{figure}

The protoplanetary disk model used in all our RHD simulations
is exactly the same as in \cite{Chrenko_etal_2017A&A...606A.114C},
including the initial and boundary conditions \citep{deValBorro_etal_2006MNRAS.370..529D}.
The parameters\footnote{A great number of the parameters listed in this section
($\nu$, $c_{\kappa}$, $A$, $T_{\star}$, $R_{\star}$, $\dot{M}_{\mathrm{F}}$, $\mathrm{Sc}$, $\epsilon_{\mathrm{p}}$, $\rho_{\mathrm{b}}$, $\alpha_{\mathrm{p}}$)
was not defined in Section~\ref{sec:model} to keep it brief.
To understand how these parameters enter the model, we refer the reader to \cite{Chrenko_etal_2017A&A...606A.114C}.}
characterizing the initial gas disk are the surface density
$\Sigma = 750(r/(1\,\mathrm{au}))^{-0.5}\,\mathrm{g}\,\mathrm{cm}^{-2}$,
kinematic viscosity $\nu=5\times10^{14}\,\mathrm{cm}^{2}\,\mathrm{s}^{-1}$,
adiabatic index $\gamma=1.4$, mean molecular weight $\mu=2.4\,\mathrm{g\,mol^{-1}}$,
vertical opacity drop $c_{\kappa}=0.6$ and disk albedo $A=0.5$.
The central star has the effective temperature $T_{\star}=4370\,\mathrm{K}$,
stellar radius $R_{\star}=1.5\,R_{\sun}$ and mass $M_{\star}=1\,M_{\sun}$.
The domain stretches from $r_{\mathrm{min}}=2.8\,\mathrm{au}$ to $r_{\mathrm{max}}=14\,\mathrm{au}$
in radius and spans the entire azimuth, having the grid resolution $1024\times1536$ (rings $\times$ sectors).

The gas disk is numerically evolved to its thermal equilibrium and
only after that, the coupled pebble disk is introduced.
Pebbles are parametrized by the radial pebble mass flux
$\dot{M}_{\mathrm{F}}=2\times10^{-4}\,M_{\earth}\,\mathrm{yr}^{-1}$ \citep{Lambrechts_Johansen_2014A&A...572A.107L},
the Schmidt number $\mathrm{Sc}=1$, coagulation efficiency $\epsilon_{\mathrm{p}}=0.5$ \citep{Lambrechts_Johansen_2014A&A...572A.107L},
bulk density $\rho_{\mathrm{b}}=1\,\mathrm{g}\,\mathrm{cm}^{-3}$ and
turbulent stirring efficiency $\alpha_{\mathrm{p}}=1\times10^{-4}$ \citep[e.g.][]{Youdin_Lithwick_2007Icar..192..588Y}.
To infer the pebble sizes, we assume the drift-limited growth regime
\citep{Birnstiel_etal_2012A&A...539A.148B,Lambrechts_Johansen_2014A&A...572A.107L}
leading to pebble sizes of several $\mathrm{cm}$.

For reference, Figure \ref{fig:aspect} shows the radial profiles of the aspect ratio $h(r)=H/r$ and temperature
$T(r)$ of the {equilibrium gas} disk.
The $h(r)$ profile has a maximum near $4\,\mathrm{au}$ where
the opacity peaks just before the sublimation of water ice \citep{Bell_Lin_1994ApJ...427..987B}. 
Therefore the heat produced by viscous heating is not easily radiated away
from this opaque region and the disk puffs up.

Near $7\,\mathrm{au}$,
the disk starts to flare because the vertical optical depth is
small enough for the incoming radiation to penetrate deeper
and heat the disk. The transition to the flared outer parts
produces a zone of convergent Type I migration for planetary
embryos \citep{Bitsch_etal_2013A&A...549A.124B,Pierens_2015MNRAS.454.2003P}.

\subsection{Nominal simulation}

\begin{figure}[]
  \centering
  \includegraphics[width=8.8cm]{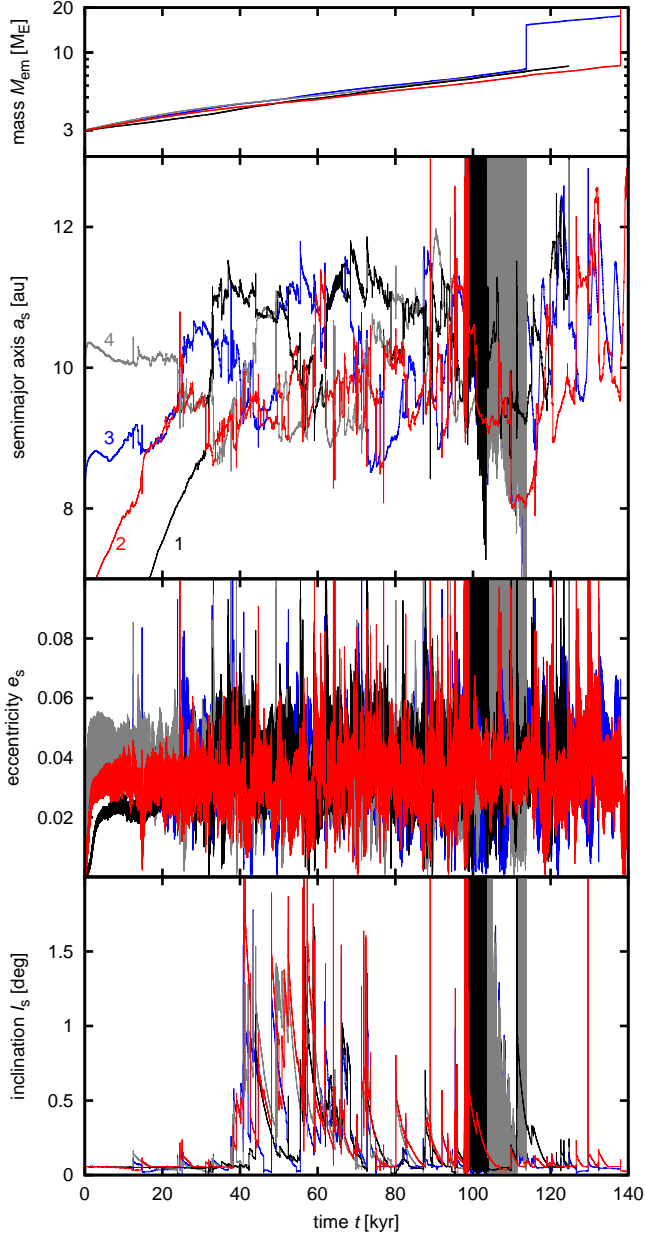}
  \caption{Temporal evolution of the embryo masses $M_{\mathrm{em}}$ (top),
  stellarcentric semimajor axes $a_{\mathrm{s}}$ (second row), eccentricities $e_{\mathrm{s}}$ (third row)
  and inclinations $I_{\mathrm{s}}$ (bottom) in the full RHD
simulation with the gas disk, pebble disk, pebble accretion and accretion
heating. Initially, there are four migrating embryos, numbered inside out.
The inclination starts at
$I_{0}=10^{-3}\,\mathrm{rad}$, which is also a value
under which the inclination damping is switched off in the model.
The strong variations of the stellarcentric Keplerian elements between $\simeq$$98\,\mathrm{kyr}$
and $\simeq$$114\,\mathrm{kyr}$ are
a consequence of binary planet formation.
A member of the binary swaps for one of the accompanying embryos three times as indicated by the
change in colors of the oscillating curves (from a narrow strip of red
to black and to gray).
The existence of the binary ends abruptly by a merger (clearly related
to the instantaneous mass increase
in the $M_{\mathrm{em}}(t)$ plot).}
  \label{fig:mtatitet}
\end{figure}

Let us now discuss and analyze our nominal simulation
in which binary planets were found to form.
Initially, we set four embryos on circular orbits with
stellarcentric semimajor axes ${a_{\mathrm{s}}}=5,6.7,8.4$ and $10.1\,\mathrm{au}$,
inclinations ${I_{\mathrm{s}}=}I_{0}=10^{-3}\,\mathrm{rad}\simeq0.057^{\circ}$
and randomized longitudes. The initial mass of each embryo is ${M_{\mathrm{em}}}=3\,{M}_{\earth}$
and their orbital separations are equal to $16$ mutual Hill radii
\begin{equation}
  R_{\mathrm{mH},ij} = \frac{a_{\mathrm{s},i}+a_{\mathrm{s},j}}{2}\left(\frac{M_{\mathrm{em},i}+M_{\mathrm{em},j}}{3M_{\star}}\right)^{1/3} \, .
  \label{eq:rmh}
\end{equation}
Embryos are numbered $1,2,3$ and $4$ from inside out.

Figure~\ref{fig:mtatitet} shows the evolution of embryos
over $140\,\mathrm{kyr}$ of the full RHD simulation with
pebble accretion and respective heating.
At first, the embryos undergo convergent
migration towards their zero-torque radius.
Without the heating torques, the embryos would concentrate
near $7\,\mathrm{au}$ thanks to the contribution
of the entropy-related corotation torque \citep{Paardekooper_Mellema_2008A&A...478..245P}
which is positive between $\simeq$$4$ to $7\,\mathrm{au}$ in this particular
disk model.

With the heating torques, however,
the zero-torque radius is shifted further out because these
torques are always positive \citep{Benitez-Llambay_etal_2015Natur.520...63B}.
Moreover, the hot-trail effect quickly excites orbital eccentricities.
Within $\simeq$$2\,\mathrm{kyr}$, the
eccentricities reach up to ${e_{\mathrm{s}}}\simeq0.02$ for the innermost embryo and ${e_{\mathrm{s}}}\simeq0.04$
for the outermost embryo.

The inclinations first remain constant near the prescribed $I_{0}$
value, with only small temporal excitations not exceeding ${I_{\mathrm{s}}}\simeq0.2^{\circ}$.
Even these initially small inclinations
are enough to modify the encounter geometries in a way that the system
becomes gradually stirred in the vertical direction.
Once the system becomes closely packed, at about $\simeq$$35\,\mathrm{kyr}$ {into the simulation},
the mutual close encounters pump the inclinations significantly, typically
to ${I_{\mathrm{s}}}\simeq1^{\circ}$ and even up to ${I_{\mathrm{s}}}\simeq2^{\circ}$.
Planets can pass above or below each other
because their vertical excursions
are comparable to (or larger than) their Hill spheres.
For example, the maximal vertical excursion
of a $5\,M_{\earth}$ embryo at $10\,\mathrm{au}$ 
is $z_{\max} \simeq 2 R_{\mathrm{H}}$ when ${I_{\mathrm{s}}}=2^{\circ}$.

Due to excited eccentricities, the embryos never form
a stable resonant chain, in accordance with \cite{Chrenko_etal_2017A&A...606A.114C}.
Moreover, the excited inclinations help the
embryos to avoid collisions and mergers for a long period of time.
Consequently, close encounters of embryos are frequent in the system.
Between $\simeq$$98\,\mathrm{kyr}$ and $\simeq$$114\,\mathrm{kyr}$,
strong unphysical oscillations of the stellarcentric orbital elements appear in Figure~\ref{fig:mtatitet}
for some of the embryos, indicating formation of gravitationally
bound binary planets\footnote{In Figure~\ref{fig:mtatitet}, we can also identify
coorbital configurations (1/1 resonances), for example for embryos $1$ and $3$ between
$62.5\,\mathrm{kyr}$ and $66\,\mathrm{kyr}$. To keep the paper
focused on binary planets, we refer an interested reader to other works
discussing formation and detectability of coorbital planets, e.g.
\cite{Laughling_Chambers_2002AJ....124..592L,Cresswell_Nelson_2008A&A...482..677C,Giuppone_etal_2012MNRAS.421..356G,Chrenko_etal_2017A&A...606A.114C,Broz_etal_2018arXiv181003385B}.
}.

\subsection{Gravitationally bound pairs of embryos}

\begin{figure}[]
  \centering
  \includegraphics[width=8.8cm]{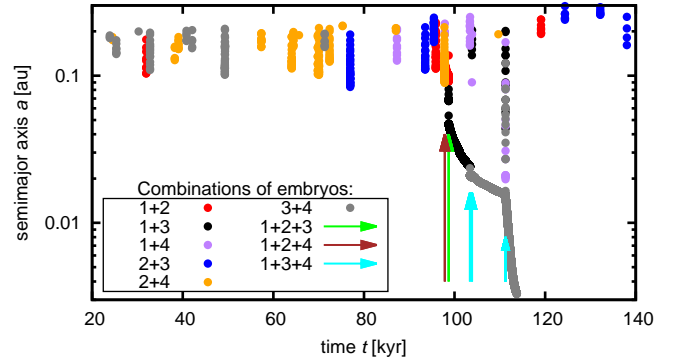}
  \caption{The record of all relative 2-body orbits
    satisfying the condition
    $a<R_{\mathrm{mH}}$ and $e<1$
    in our nominal simulation. The filled circles mark the instantaneous
    semimajor axis and each pair of embryos is distinguished by color.
    The arrows are used to indicate when a gravitationally bound pair
    participates in a 3-body encounter.
  }
  \label{fig:encounter_events}
\end{figure}

To identify the events related to the binary formation in the
simulation described above,
we computed orbits of the relative motion among all possible pairs
of embryos and selected those with
$a<R_{\mathrm{mH}},e<1$. The results are shown
in Figure~\ref{fig:encounter_events}.
Throughout the simulation, we found $65$
time intervals during which at least
two embryos are captured on a mutual elliptic orbit,
changing their relative orbital energy from
initially positive to negative (and back to positive
when the capture terminates).
Subsequently, we also scanned the sample of bound pairs 
and looked for cases when a third embryo has its distance from a
pair $d<R_{\mathrm{mH}}$ to identify 3-body (pair-embryo) encounters.
These are highlighted in Figure~\ref{fig:encounter_events} by arrows.

Analyzing the lifetime of the bound pairs, we found that
most of them dissolve before finishing one stellarcentric orbit.
However, we found a single case when several
binary configurations existed consecutively
for a prolonged period of time between $\simeq$$98\,\mathrm{kyr}$ and $\simeq$$114\,\mathrm{kyr}$.
This time interval is bordered by 3-body encounters
which usually cause the binary separation to drop.

In summary, bound pairs can
form in 2-body encounters but they quickly dissolve
unless they undergo a 3-body encounter with one of the
remaining embryos. The latter process is known as binary hardening
\citep[e.g.][]{Hills_1975AJ.....80..809H,Hills_1990AJ.....99..979H,Goldreich_etal_2002Natur.420..643G,Astakhov_etal_2005MNRAS.360..401A}
and it occurs when an external perturber removes energy
from a binary system which then becomes more tightly bound.
To distinguish among two types of events contributing
to the formation of bound pairs,
we call those formed in 2-body encounters
{\em transient} binaries because of their typically short dynamical lifetime.

The transient binaries have been a subject of many different studies,
for example, in the 3-body Hill problem \citep[e.g.][]{Simo_Stuchi_2000PhyD..140....1S,Astakhov_etal_2005MNRAS.360..401A}.
Their formation is possible because the orbital
energy of two bodies in no longer conserved when additional
perturbers (e.g. the central star) are present
\citep[e.g.][]{Cordeiro_etal_1999AJ....117.1634C,Araujo_etal_2008MNRAS.391..675A}.
Our system is of course more complicated because
additional gravitational perturbations arise
from the gas disk. We will demonstrate in
Section~\ref{sec:transient_formation} that the gas indeed
facilitates formation of transients.

One last question we address here is whether or not
the occurrence of bound pairs in Figure~\ref{fig:encounter_events}
is related to the vertically stirred orbits of embryos.
To find an answer, we looked for bound pairs
(with $a<R_{\mathrm{mH}},e<1$) in one of our previous simulations
reported in \cite{Chrenko_etal_2017A&A...606A.114C} (dubbed Case III),
where the inclinations were damped in the standard way \citep{Tanaka_Ward_2004ApJ...602..388T}.
We found only $6$ bound pairs (all transients) compared to $65$ cases
in our nominal simulation presented here. Therefore,
the excited inclinations importantly change the outcomes of embryo
encounters in the gas disk and help to form transients.

\subsection{Transient binary formation in a 2-body encounter} 
\label{sec:transient_formation}

\begin{figure*}[]
  \centering
  \begin{tabular}{ccc}
  (a) & (b) & (c) \\
  \includegraphics[width=5.86cm]{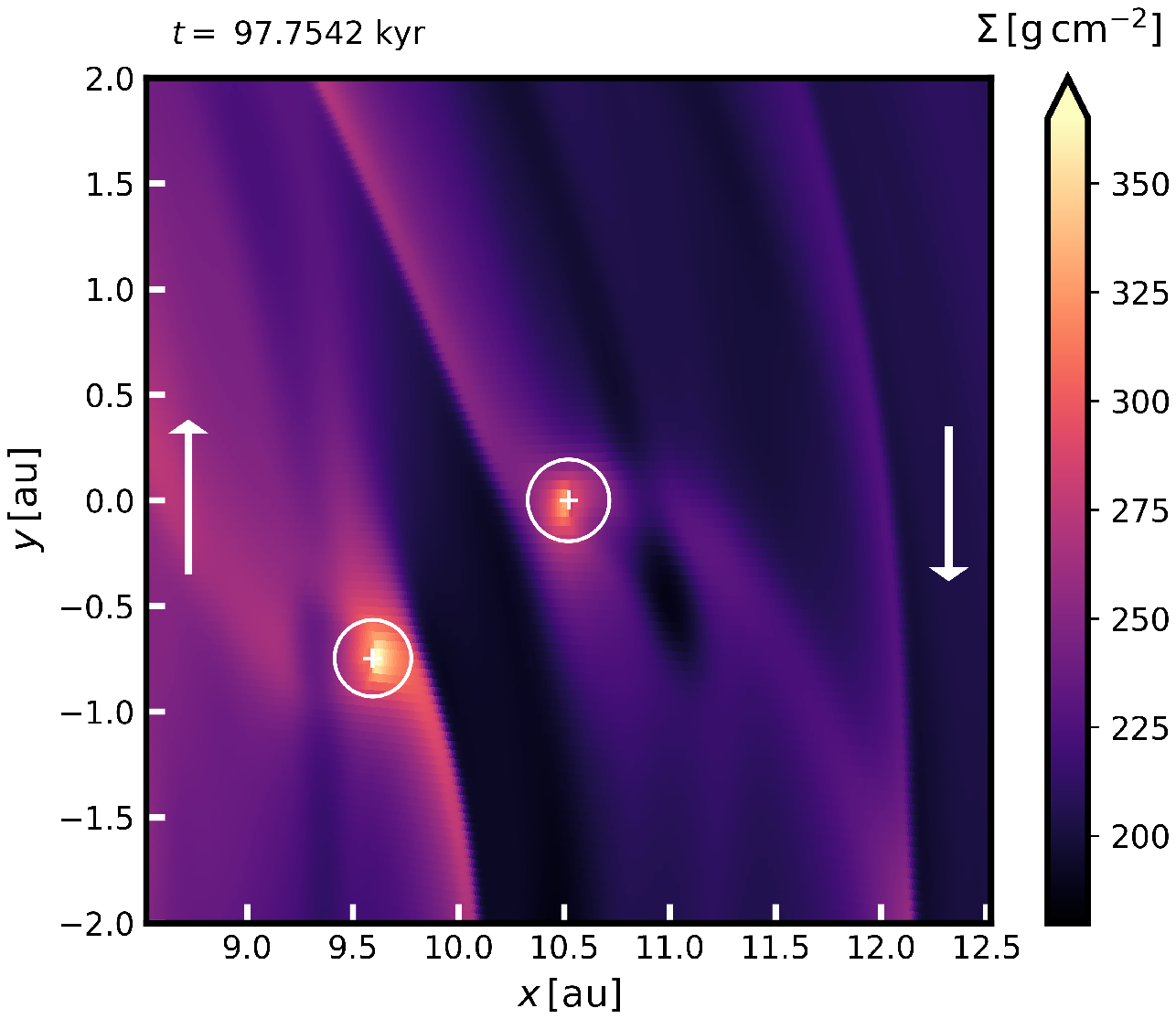} &
  \includegraphics[width=5.86cm]{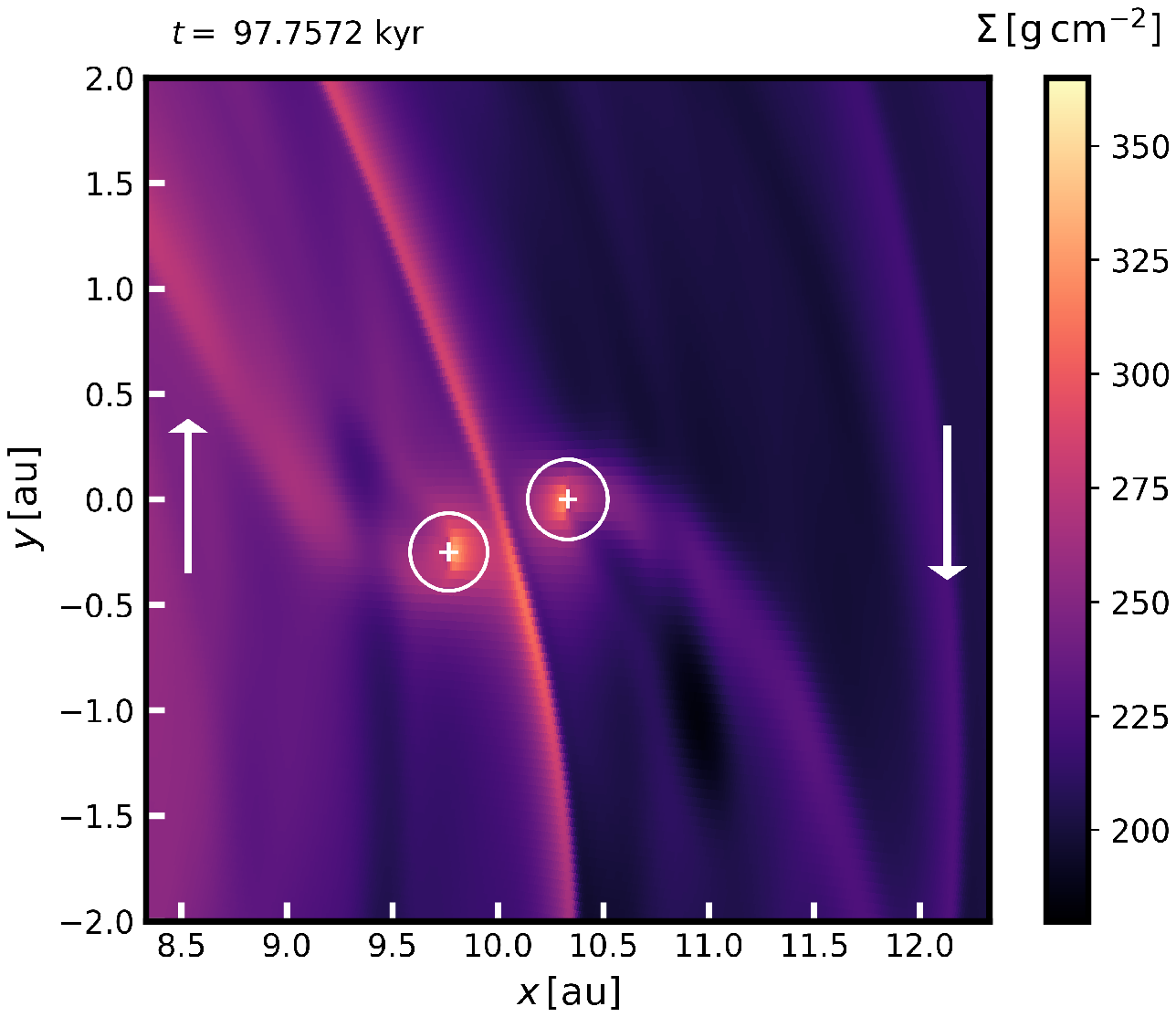} &
  \includegraphics[width=5.86cm]{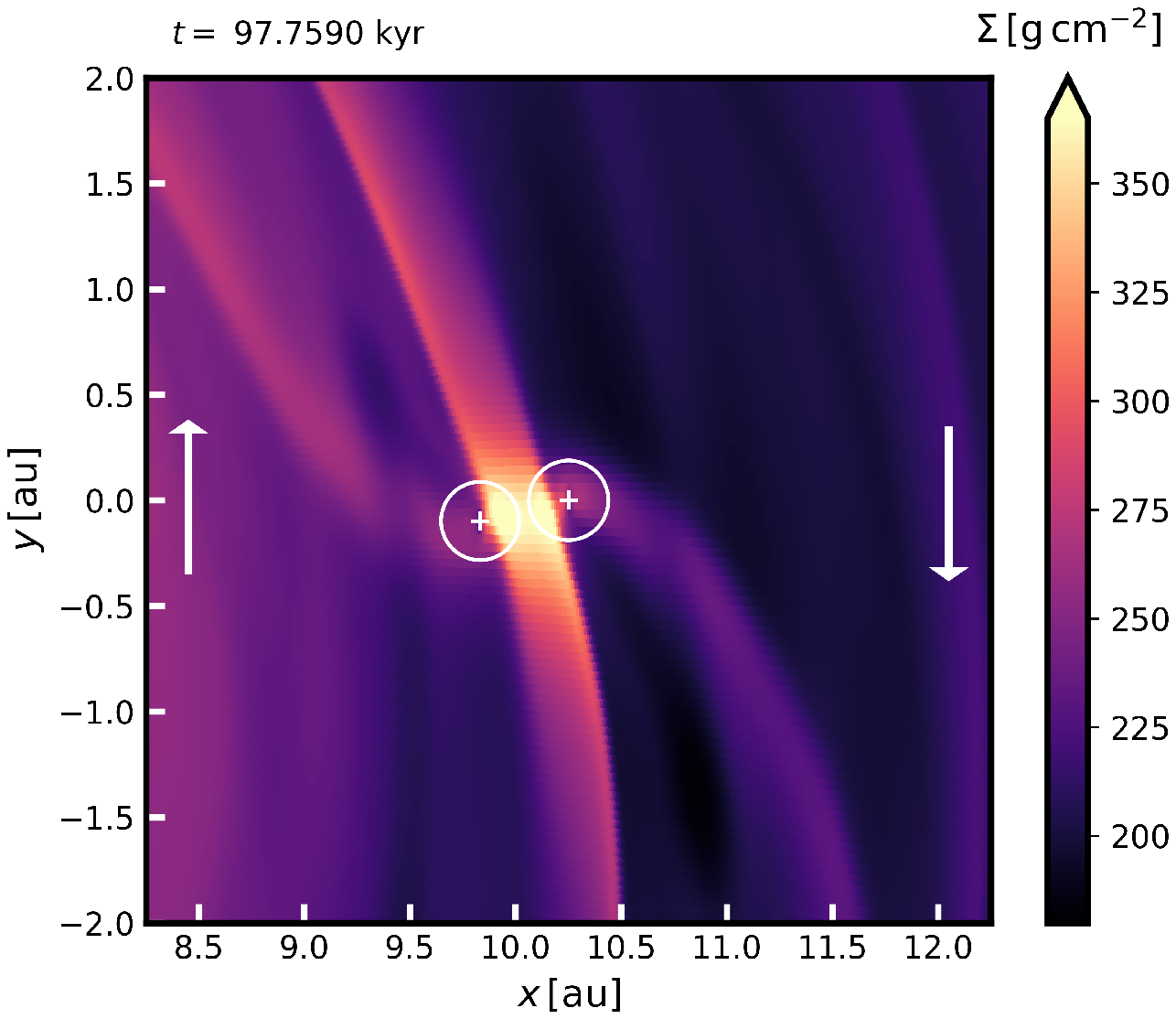} \\
  \includegraphics[width=5.86cm]{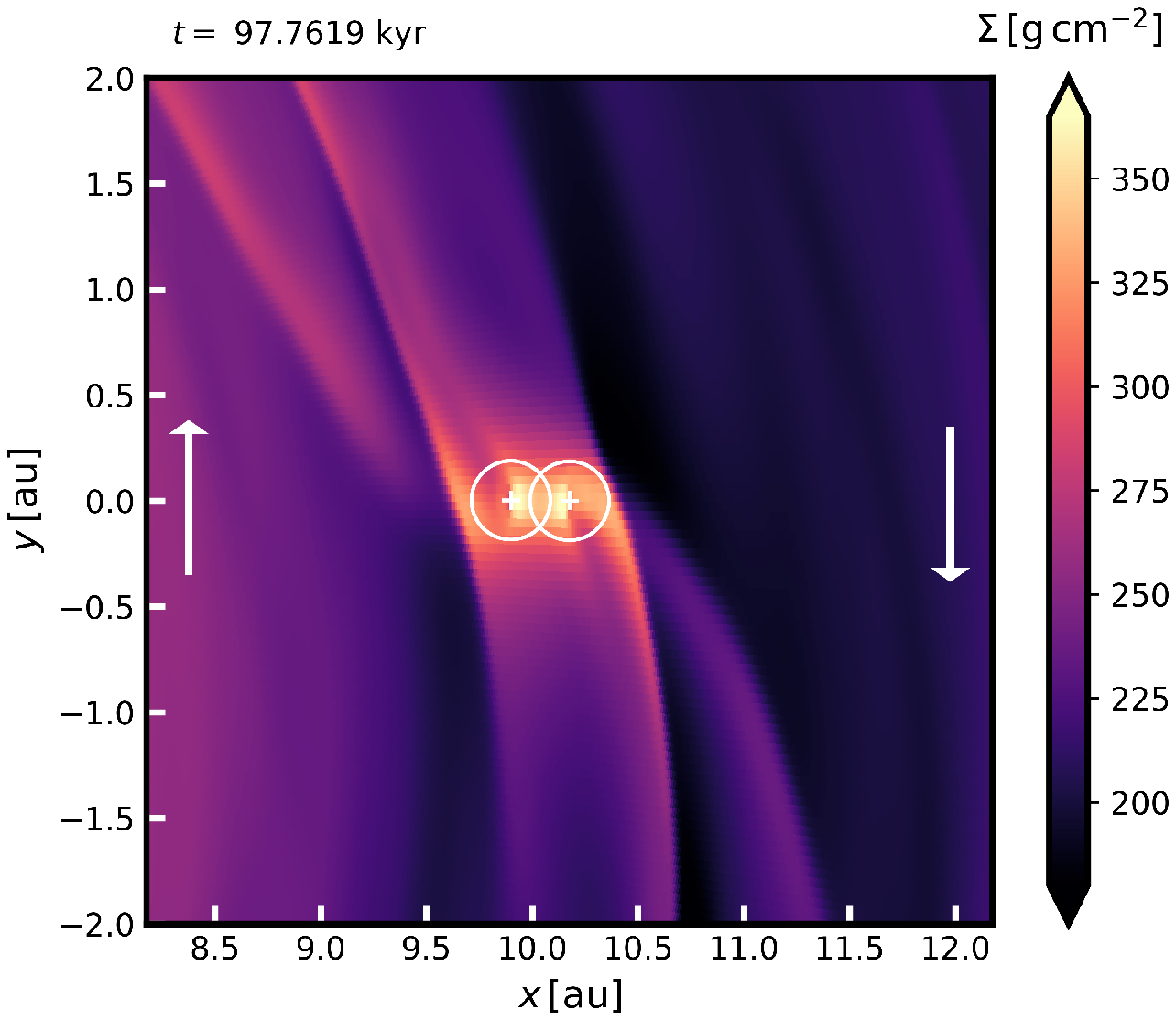} &
  \includegraphics[width=5.86cm]{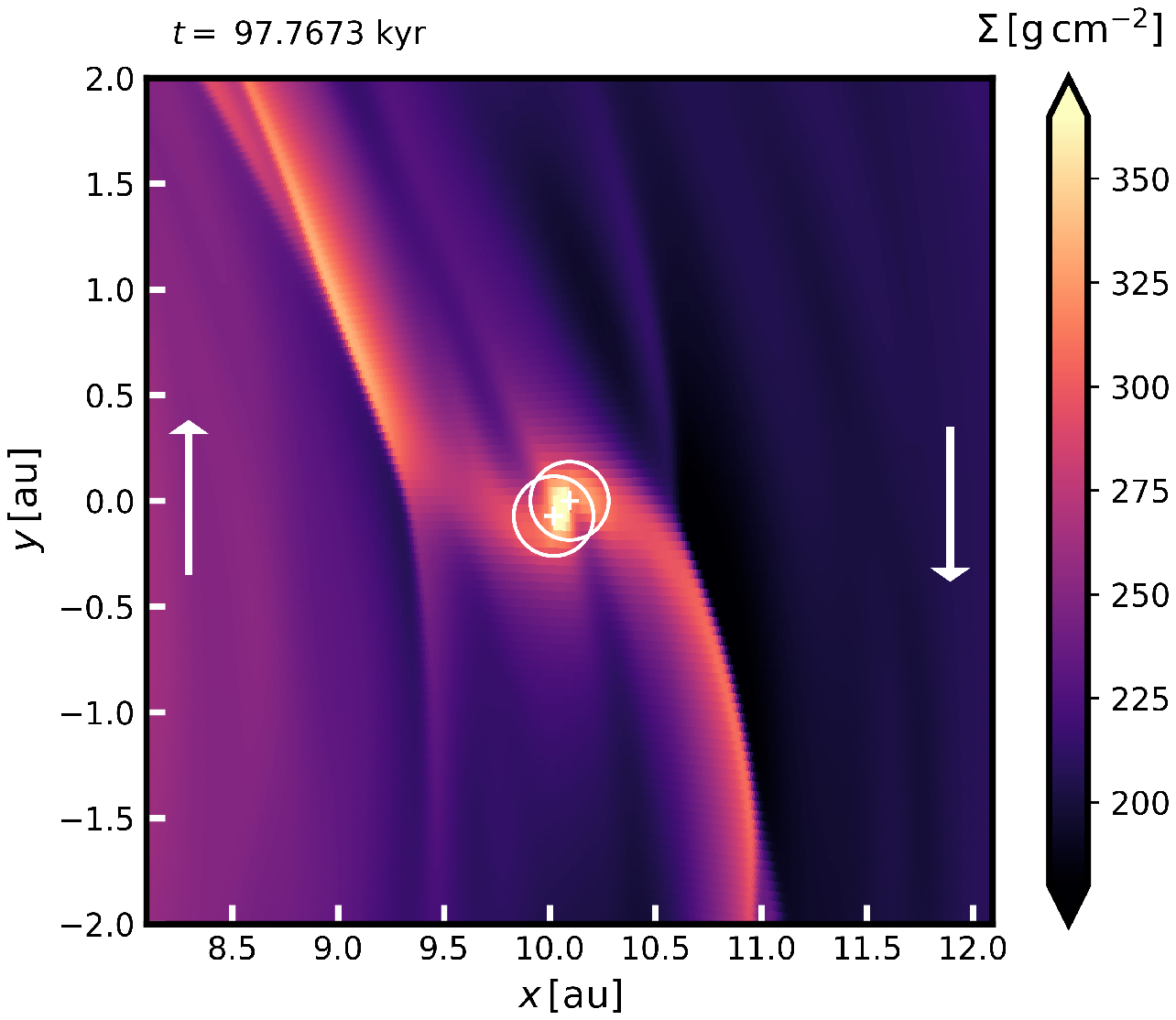} &
  \includegraphics[width=5.86cm]{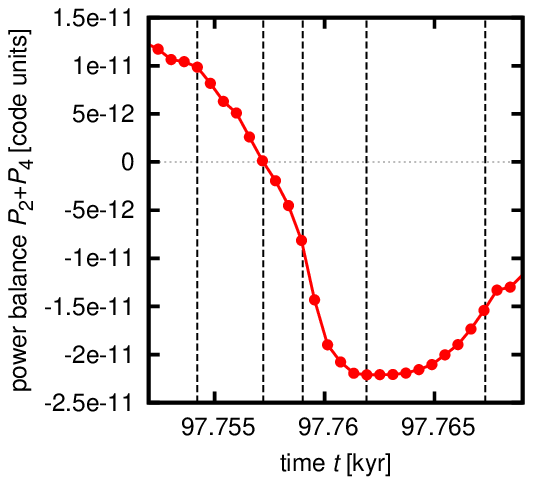} \\
  (d) & (e) & (f) \\
  \end{tabular}
  \caption{
    Formation of a transient binary during a 2-body encounter of embryos 2 and 4.
    The first five panels (a)--(e) 
    show the evolution of the perturbed surface gas density $\Sigma$.
    Locations of the embryos are marked by white crosses, their Hill spheres are indicated
    by white circles and white arrows show the direction of the Keplerian shear in the reference
    frame corotating with the embryo 2 which is placed at the center of each plot.
    The individual panels are labeled with the simulation time $t$. The final panel (f)
    shows the evolution of the total power of the gravitational forces exerted by the disk onto
    embryos 2 and 4. The vertical dashed lines mark the simulation time
    corresponding to the snapshots of $\Sigma$ in panels (a)--(e).
    An animation of the formation of the transient binary is
        available in the online Journal and covers
	$t\simeq97.73$--$97.8\,\mathrm{kyr}$ in simulation time.
  }
  \label{fig:transient_formation}
\end{figure*}

Here we investigate formation of a transient pair of embryos $2$ and $4$ which precedes
the binary hardening events in our nominal simulation.
The pair forms in a 2-body interaction at $t\simeq97.76\,\mathrm{kyr}$. 
To see whether the embryo-disk interactions assist in the process,
we show in Figure~\ref{fig:transient_formation} the evolution of the
perturbed gas surface density $\Sigma$ during the encounter.

Before the encounter (panel (a)), the usual structures can be seen in the disk.
The hot (underdense) trail of the outer embryo 2 can be seen as a dark
oval spot in the bottom right quadrant (it moves down and gets larger in time).
The hot trail of the inner embryo 4 is less prominent and looks
as an underdense gap attached to the embryo from inside.

As the embryos approach in panel (b), the outer spiral arm of the inner
embryo 4 and the inner spiral arm of the outer embryo 2 join and overlap.
The overlap forms a strong density wave positioned between the embryos.
The overdensity increases as part of the wave becomes
trapped between the Hill spheres of both embryos in panel (c).
From panel (c) to (d), the embryos cross this shared density wave.

In panel (d), the embryos are so close to each other that they effectively
act on the disk as a single mass and the previously shared spiral arm
splits into an inner and outer component with a small pitch angle.
There are two more spirals with a larger pitch angle which are leftovers
of the initial wakes launched by the embryos. In panel (e), all
spirals blend into a single pair of arms. The embryos
enter one another's Hill sphere between panels (d) and (e) through
the vicinity of the Lagrange points $\mathrm{L}_{1}$ for the outer
and $\mathrm{L}_{2}$ for the inner embryo, respectively \citep{Astakhov_etal_2005MNRAS.360..401A}.
The embryos are captured on a prograde binary orbit (in panel (e), the embryo 4 orbits
the central embryo 2 counterclockwise).

The spiral arm crossing which appears during the encounter
is known to produce strong damping effects onto the embryos
\citep{Papaloizou_Larwood_2000MNRAS.315..823P}. It is thus 
likely that the gas supports the gravitational capture by dissipating the orbital
energy. To quantify this effect, we measured the total gravitational
force $\vec{F}_{\mathrm{g},i}$ exerted by the disk onto each embryo and 
we calculated the mechanical power
\begin{equation}
  P_{i} = \int\limits_{\mathrm{disk}}\vec{v}_{i}\cdot \vec{F}_{\mathrm{g},i}\mathrm{d}S \, ,
  \label{eq:power}
\end{equation}
where $\vec{v}_{i}$ is the velocity vector of the embryo and the integral
goes over the entire disk. $P_{i}$ directly
determines the rate of change of the orbital energy $\dot{E}_{i}$ of each
embryo \citep[e.g.][]{Cresswell_etal_2007A&A...473..329C}.

Panel (f) of Figure~\ref{fig:transient_formation} shows the total balance of the energy
subtraction (or addition) for the embryos $2$ and $4$, $P=P_{2}+P_{4}$. The
energy is transfered to the disk when $P<0$ and vice versa. It is obvious
that throughout the closest approach (panels (b)--(e)), the orbital energy
of the embryos dissipates and thus the influence of the gas disk 
on formation of transients is confirmed.

\subsection{Binary planet hardening in 3-body encounters}

\begin{figure*}[]
  \centering
  \begin{tabular}{cc}
  \includegraphics[width=8.8cm]{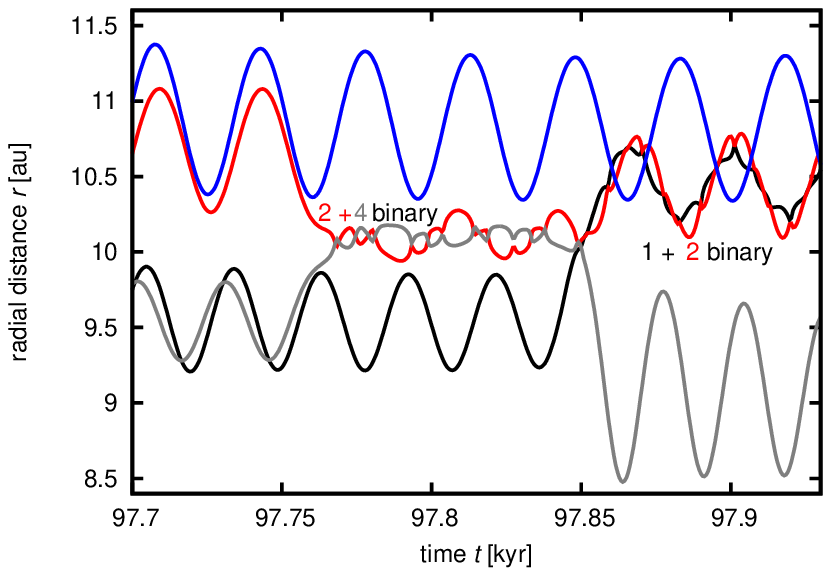} &
  \includegraphics[width=8.8cm]{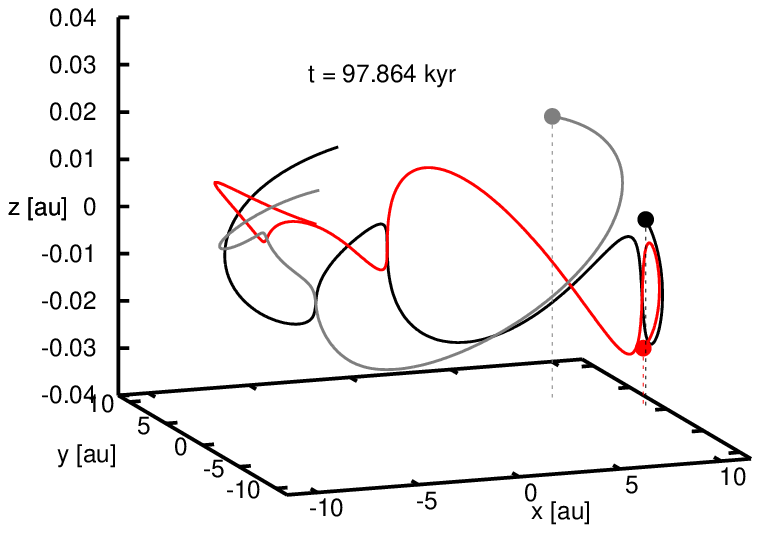} \\
  \includegraphics[width=8.8cm]{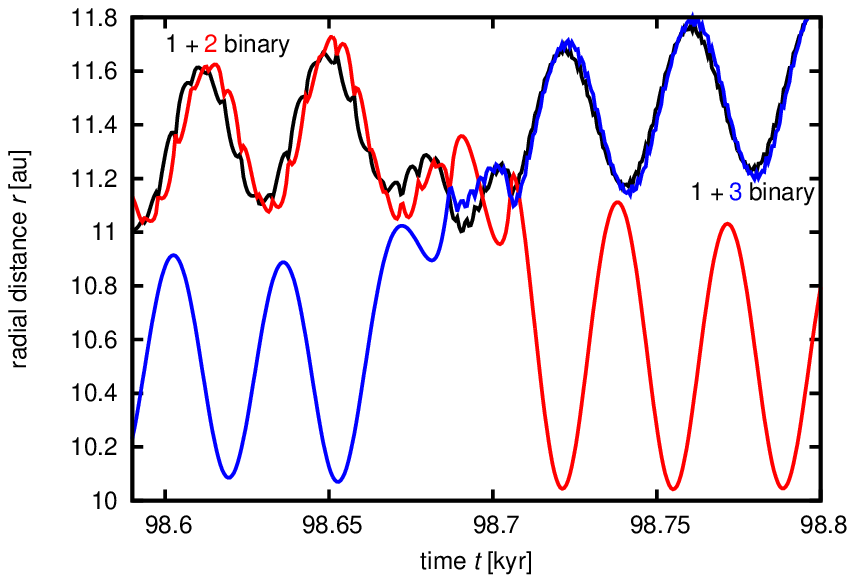} &
  \includegraphics[width=8.8cm]{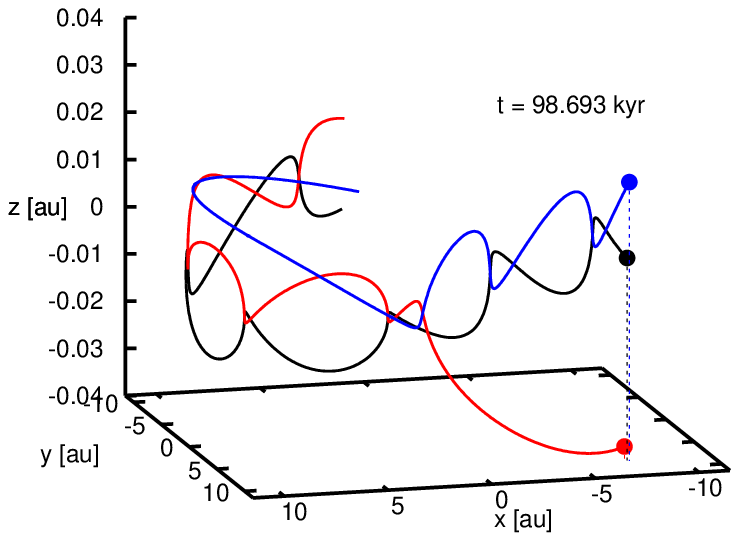} \\
  \includegraphics[width=8.8cm]{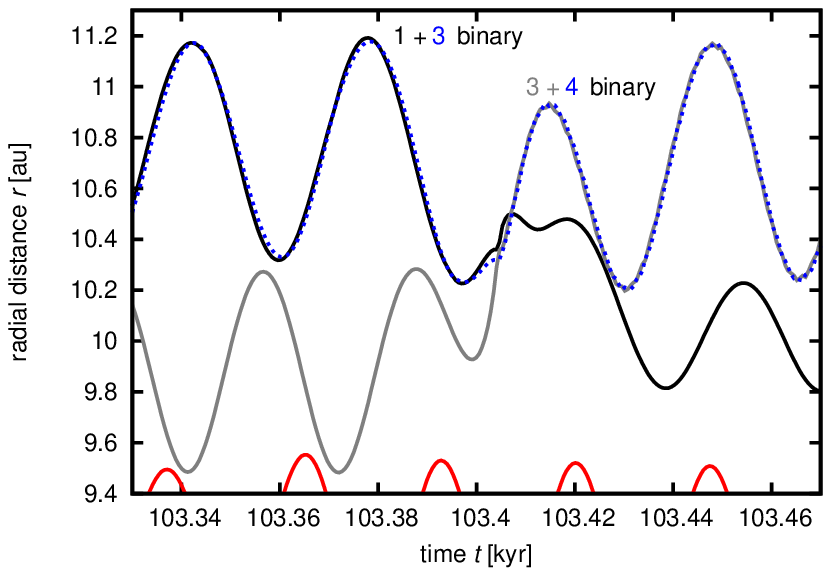} &
  \includegraphics[width=8.8cm]{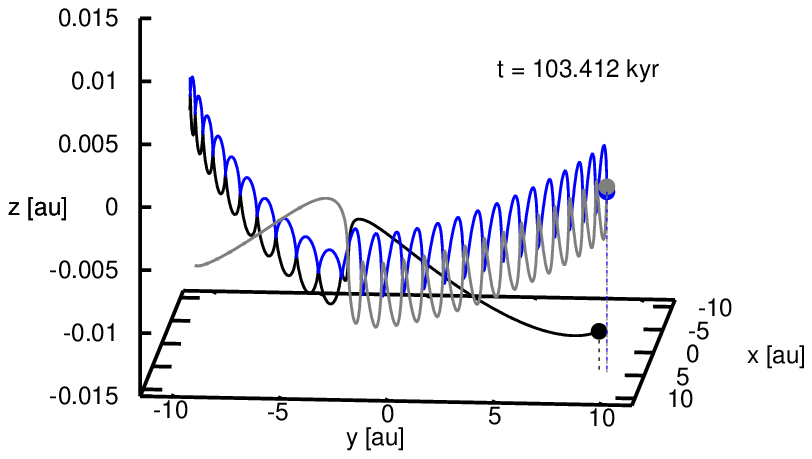} \\
  \end{tabular}
  \caption{Details of important 3-body encounters occurring in the nominal
    simulation, leading to reconfiguration and hardening of the binary
    planet. Each row shows a different encounter (and the first row also
    shows the transient binary of embryos 2+4 preceding the 3-body
    encounters). We plot the temporal evolution of the radial distance $r$
    (left column) and also the orbital evolution in 3D Cartesian space
    (right column) in a short time interval around the respective
    encounters. The numbering and coloring of the bodies is the same as in
    Figure~\ref{fig:mtatitet}. Filled circles in the right column mark the
    positions of embryos at the simulation time $t$ given by the labels.
    An animated version of the top row of Figure~\ref{fig:3body_events}
        is available in the online Journal. The animation spans
	$t\simeq97.7$--$97.92\,\mathrm{kyr}$ in
        simulation time, displaying the same tracks as in the static figure.
    }
  \label{fig:3body_events}
\end{figure*}

The transient pair of embryos $2$ and $4$ does not dissolve,
instead it is further stabilized in 3-body encounters with the remaining
embryos. Figure~\ref{fig:3body_events} shows these encounters
in detail. First, the transient pair encounters embryo $1$
at $t\simeq97.85\,\mathrm{kyr}$. 
During this encounter, one component of the binary (embryo $4$)
becomes unbound and is deflected away but the incoming embryo $1$
takes its place in an exchange reaction so the binary does not cease to exist.

Similar situation repeats at about $\simeq$$98.69\,\mathrm{kyr}$
when the configuration $1+2$ changes to $1+3$ and 
at $\simeq$$103.4\,\mathrm{kyr}$
when the configuration $1+3$ changes to $3+4$.
Figure~\ref{fig:3body_events} also reveals the binary becomes
hardened with each encounter since the overlap between the $r(t)$ curves for
the binary components becomes tighter.

\begin{figure}[]
  \centering
  \includegraphics[width=8.8cm]{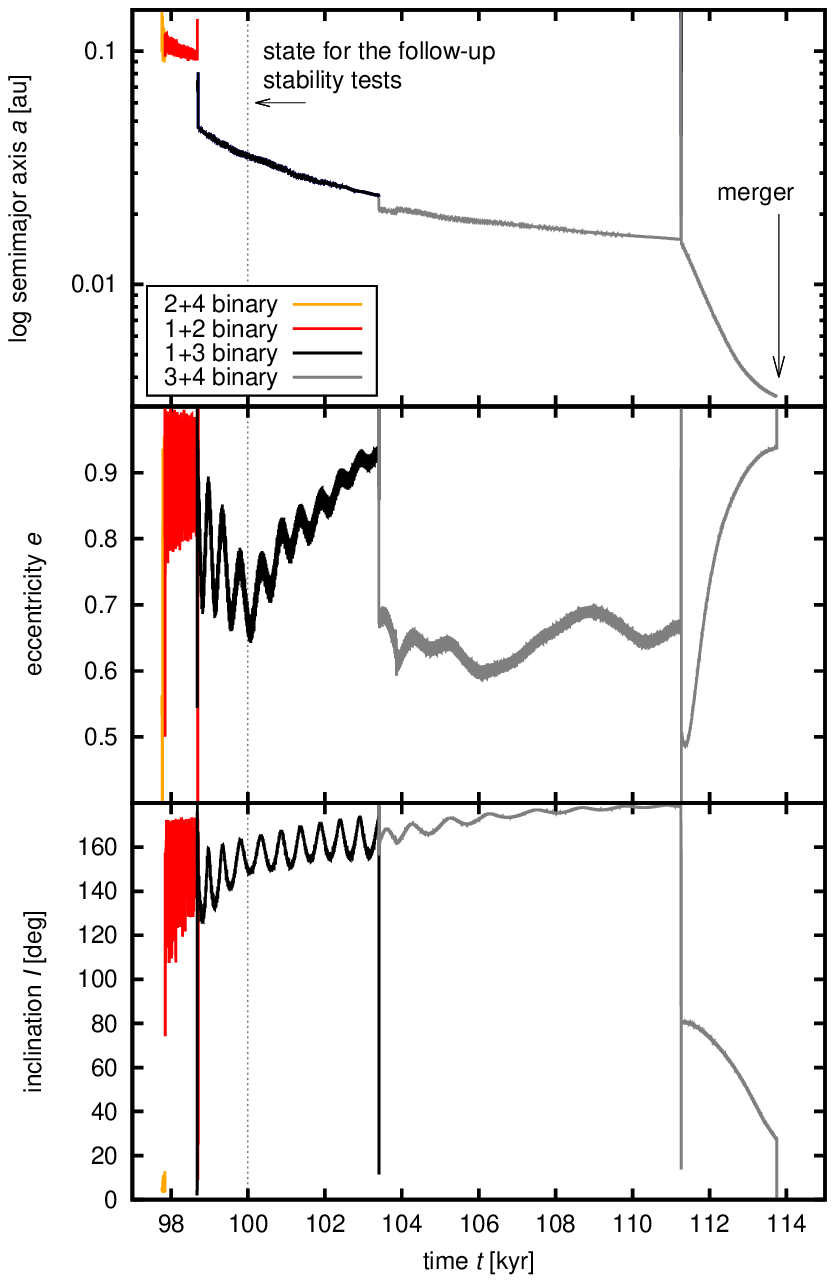}
  \caption{The temporal evolution of the orbital elements
    of the binary planet, namely the semimajor axis $a$ (top),
    eccentricity $e$ (middle) and inclination $I$ (bottom).
    The color of the curves changes if different embryos
    become locked in the binary configuration during a 3-body
    encounter (see Figure~\ref{fig:3body_events} for reference).
    The encounters can be recognized as sudden spikes
    and they typically lead to a sudden decrease
    of $a$ and $e$.
    In the course of time, the binary becomes bound
    more tightly. The existence of the binary
    ends with a merger, marked by an arrow.}
  \label{fig:atetit_binary}
\end{figure}

Even clearer indication of the binary hardening is provided by Figure~\ref{fig:atetit_binary}
where we plot the temporal evolution of the orbital elements
of the binary planet. The color
of the curves changes each time there
is a change in the composition of the binary.
One can further notice
the exchange interactions produce sudden
decreases of $a$ and also of $e$.
Between the exchange interactions, $a$ smoothly
decreases whereas $e$ generally increases in an oscillatory manner.
We will describe these variations later.

Considering the binary planet inclination,
the transient binary $2+4$ forms with a prograde orbit
and a relatively low inclination.
During the first 3-body encounter, the binary 
is reconfigured to retrograde orbit with the inclination oscillating
between $100^{\circ}$ and $170^{\circ}$.
The inclination then slowly evolves towards $180^{\circ}$ regardless of the
swap encounters which only diminish 
the oscillation amplitude.

The binary planet does not survive to the end of our simulation. 
At $\simeq$$111.26\,\mathrm{kyr}$, it undergoes a 3-body exchange
interaction with the embryo $1$ during which their
Hill spheres overlap for a prolonged period of time
(see the spike in Figure~\ref{fig:encounter_events} at $\simeq$$111.26\,\mathrm{kyr}$).
Consequently, the binary inclination is flipped from
the retrograde configuration to $I=80^{\circ}$.
In this configuration, the binary undergoes fast decrease of $a$,
accompanied by an equally fast increase of $e$. Consequently,
the binary planet ends its life in a merger into a single body.
  
\subsection{Binary planet evolution without perturbing embryos}
\label{sec:stab_isol}

The lifetime of the hardened binary in our nominal simulation is 
long enough ($\sim$$10^{4}\,\mathrm{yr}$) to be interesting.
There are two basic questions that we shall now address.
First, what is the evolution of such a binary
if the surrounding embryos and their perturbations are ignored?
And second, what causes the changes of the binary orbital elements
between the 3-body encounters in Figure~\ref{fig:atetit_binary}?

\begin{figure}[]
  \centering
  \includegraphics[width=8.8cm]{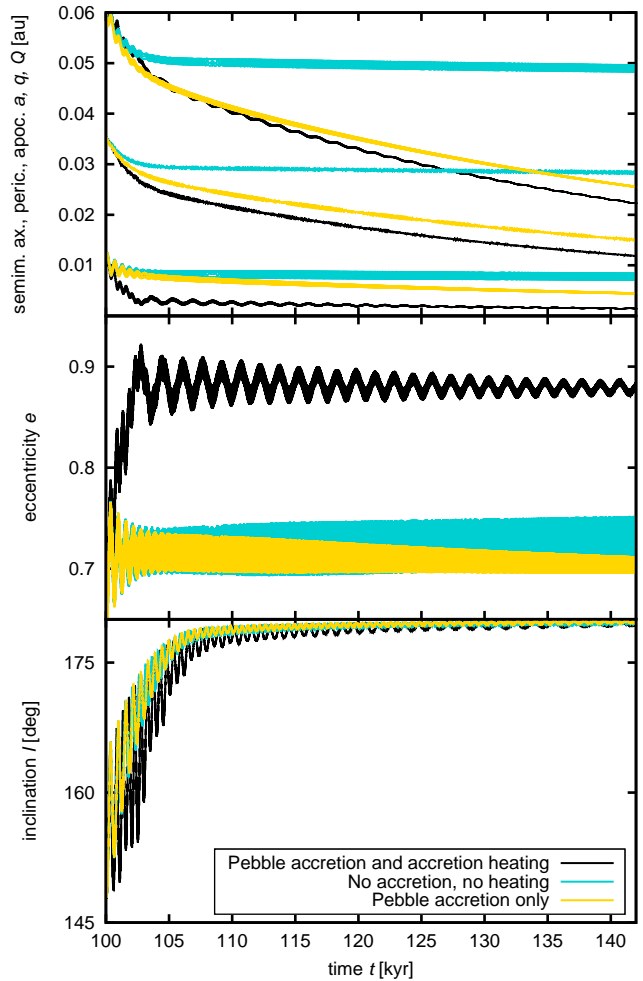}
  \caption{Stability test starting from the binary planet configuration corresponding
    to $t=100\,\mathrm{kyr}$ in Figure~\ref{fig:atetit_binary}. In this test,
    the surrounding planetary embryos are removed. Therefore we study
    evolution of an isolated binary driven only by its interactions with the disk
    and not by close encounters.
    Three cases are shown: One with the complete RHD and two-fluid part of the model (black curve),
    one with pebble accretion but without accretion heating (gold curve),
    and one without both pebble accretion and heating (turquoise curve).
    Pebble accretion (i.e. mass growth) of the embryos generally causes
    the semimajor axis decrease. Accretion heating, on the other hand,
    is capable of pumping the eccentricity above the initial value.
  }
  \label{fig:atetit_ISOL_binary}
\end{figure}

To answer these questions, we discard the non-binary embryos
and restart the simulation from the configuration
of the binary planet\footnote{{The binary elements at the moment of
restart are $a\simeq0.035\,\mathrm{au}$, $e\simeq0.66$, $I\simeq2.6\,\mathrm{rad}$.}},
gas and pebbles corresponding to $t=100\,\mathrm{kyr}$.
Three models are numerically evolved for $45\,\mathrm{kyr}$.
The first one has the same setup as the initial simulation
(apart from the ignored non-binary embryos). In the second one,
the accretion heating is disabled but the mass of the binary components
can still grow by pebble accretion. In the third one,
we again switch off accretion heating and discard
the pebble disk; the binary mass therefore remains constant.

The orbital evolution of the binary in these three cases is 
shown in Figure~\ref{fig:atetit_ISOL_binary}.
The inclination evolution is more or less the same, regardless
of the model, and converges toward a fully retrograde configuration.
The semimajor axis decreases as a consequence
of pebble accretion which transports the linear momentum
and mass onto the binary components, thus changing
their orbital angular momentum.
It is worth noting that if the pebble accretion and accretion heating
are ignored, the isolated binary planet evolving
in the radiative disk exhibits only minor orbital changes
(once it adjusts to the removal of the surrounding embryos at the beginning
of the restart).

The eccentricity substantially changes only in the model with accretion
heating, otherwise it oscillates around its initial value
or exhibits a slow secular variation.
In other words, the hot-trail effect is important not only
for exciting the eccentricities of individual embryos
before the encounter phase, but it is also responsible for pumping
the eccentricity of the binary up to an asymptotic value $e\simeq0.75$.

These findings justify our incorporation of pebble accretion and accretion
heating into the model because both phenomena affect the rate of change
of the binary orbital elements. Pebble accretion diminishes the semimajor axis
and accretion heating excites the eccentricity.

\subsection{Binary planet evolution in the disk-free phase}
\label{sec:nbody_models}

\begin{figure}[]
  \centering
  \includegraphics[width=8.8cm]{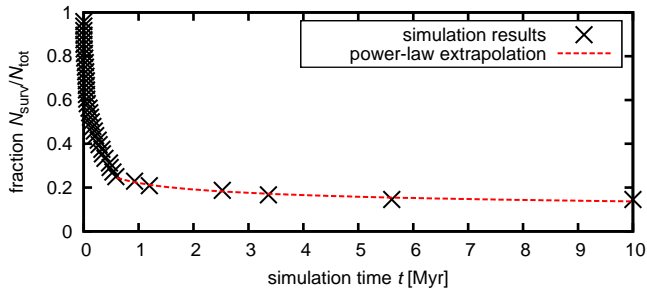}
  \caption{
    The fraction of
    N-body integrations in which the binary planets survives
    until simulation time $t$. The crosses mark the results
    of our integrations, the dashed curve is a power-law
    fit which is used as an extrapolation for $t>10\,\mathrm{Myr}$.
  }
  \label{fig:lifetime}
\end{figure}

\begin{figure*}[]
  \centering
  \includegraphics[width=18.0cm]{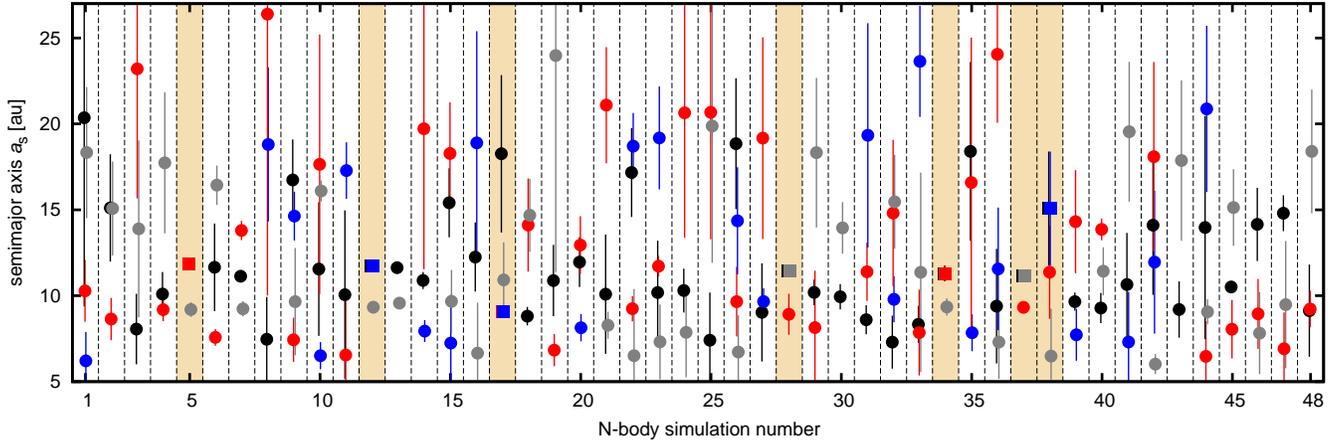}
  \caption{
      Orbital distribution of the planetary systems $10\,\mathrm{Myr}$
      after the disk removal. The horizontal axis shows the reference
      number of N-body simulations, individual cases are separated
      by the vertical dashed lines. The vertical axis shows the
      stellarcentric semimajor axis of planets displayed with symbols (the coloring
      is the same as in the previous figures). Single planets are marked
      by filled circles, binary planets are distinguished by filled squares
      (squares correspond to $a_{\mathrm{s}}$ of the binary barycenter).
      The systems which preserved the binary are also highlighted using the yellow background.
      The vertical bars of each symbol indicate the span of orbits from pericenter to apocenter (large
      bars correspond to eccentric orbits and vice versa).
  }
  \label{fig:ae_nbodyconfigs}
\end{figure*}

When protoplanetary disks undergo dispersal due to photoevaporation,
the emerging planetary systems may become unstable \citep[e.g.][]{Lega_etal_2013MNRAS.431.3494L}.
Here we test whether the hardened binary planet could survive
the gas removal phase and the subsequent orbital instabilities.
Since the binary undergoes 3-body encounters during the disk
phase, they can be expected also after the disk removal.

To investigate the evolution after the photoevaporation,
we remove the fluid part of the model (i.e. gas and pebbles) instantly and 
continue with a pure N-body simulation. The orbits
are integrated for additional $10\,\mathrm{Myr}$.
To account for the chaotic nature of an N-body system
with close encounters, we extract $48$ orbital configurations
of the embryos from between $\simeq$$99.7\,\mathrm{kyr}$ and $\simeq$$102.4\,\mathrm{kyr}$
of the nominal simulation and use them as the initial
conditions for $48$ independent integrations.

Our aim is to quantify the binary
  planet survival rate. Figure~\ref{fig:lifetime} shows
  the evolution of the fraction $N_{\mathrm{surv}}/N_{\mathrm{tot}}$,
  where $N_{\mathrm{surv}}$ is the number of N-body
  systems still containing a binary planet at simulation time $t$
  and $N_{\mathrm{tot}}$ is the total number of systems (48).
  The dependence exhibits a steep decrease ---
  the binary dissolves before $0.1\,\mathrm{Myr}$ of evolution
  in $\simeq$$45\,\%$ of cases and before $1\,\mathrm{Myr}$ in
  $\simeq$$76\,\%$ of cases.
  However, the trend for $t\geq1\,\mathrm{Myr}$ becomes rather flat.
  The binary planet survives the whole integration timespan in $15\,\%$
  of our runs.

We estimate the fraction of planetary systems $f_{\mathrm{surv}}$
in which the binary can survive the orbital instabilities.
An estimate for young systems can be readily done by taking
the final fraction of our integrations conserving the binary,
yielding $f_{\mathrm{surv},10\,\mathrm{Myr}}\simeq15\%$.
To make an estimate for older systems, we performed a power-law extrapolation
of the flat tail of the distribution in Figure~\ref{fig:lifetime}.
The resulting extrapolation $0.22(t/(1\,\mathrm{Myr}))^{-0.21}$
yields $f_{\mathrm{surv},4.5\,\mathrm{Gyr}}\simeq4\%$
for $t=4.5\,\mathrm{Gyr}$ (i.e. comparable to the Solar System age).

Let us briefly discuss the orbital architecture
of the individual systems at the end of our integrations
(Figure~\ref{fig:ae_nbodyconfigs}). Focusing
on the systems which retained the binary, they
can be divided into two classes. The more common first class
comprises five systems (simulation number $5,12,28,34,37$)
in which one of the binary components undergoes
an early collision (at about $\simeq0.1\,\mathrm{Myr}$)
with one of the remaining embryos
while maintaining the binary configuration.
The collision reduces the multiplicity of the system
and changes the mass ratio of the binary components from $1/1$ to $2/1$.
The stability of such systems is obvious from
Figure~\ref{fig:ae_nbodyconfigs} because eccentricities
are only marginally excited and the planets do not
undergo orbital crossings.

The less frequent second class of orbital architectures
(simulation number $17$ and $38$) includes systems in which
no collision occurred yet the binary managed to survive
close encounters. Orbits of the single embryos
in these simulations are moderately eccentric and have
orbital crossings with the binary. It is very likely
that these systems would reconfigure and the binary
  would dissolve
  when integrating for $t>10\,\mathrm{Myr}$.
We however
believe that these two cases are accounted
for by the power-law extrapolation when estimating
$f_{\mathrm{surv}}$ for old systems.

\subsection{Formation efficiency}
\label{sec:formation_efficiency}

\begin{figure}[]
  \centering
  \includegraphics[width=8.8cm]{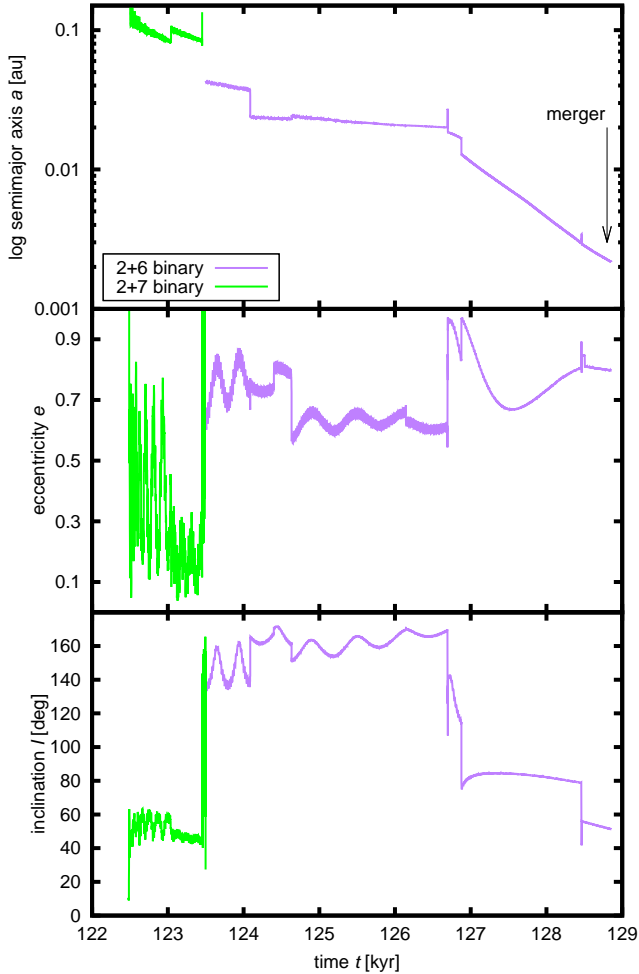}
  \caption{As Figure~\ref{fig:atetit_binary}, but for the binary configurations
  found in our additional simulation IV with 8 embryos (initial $M_{\mathrm{em}}=1.7\,\mathrm{M_{\earth}}$).
  The binary shown here undergoes similar evolution to that in Figure~\ref{fig:atetit_binary}.}
  \label{fig:verif_binaries}
\end{figure}

So far the analysis was based on our nominal simulation.
Although a broader study of the parametric space is difficult using RHD
simulations, it is important to quantify how common it is for transients
to become hardened and stabilized in 3-body encounters.
Also, it is desirable to test if the result is sensitive to the choice
of initial separations, embryo masses and embryo multiplicity.

We perform four additional full RHD simulations 
in which we vary the initial conditions for embryos
(the disk remains the same as in Section~\ref{sec:diskmodel}).
Embryos start with different random
longitudes and inclinations are $I_{\mathrm{s}}=I_{0}$.
Two simulations (denoted I and II) are run with 4 embryos,
each having $M_{\mathrm{em}} = 3\,\mathrm{M_{\earth}}$ again, but
the innermost embryo is initially placed at $a_{\mathrm{s}}=6\,\mathrm{au}$
and the remaining ones are spaced by $10$ mutual Hill radii.
Two simulations (denoted III and IV)
include 8 embryos with $M_{\mathrm{em}} = 1.7\,\mathrm{M_{\earth}}$,
the inner one being placed at $a_{\mathrm{s}}=5\,\mathrm{au}$
and the others having initial separations of $8$ mutual Hill radii.

The simulations I and II cover $120\,\mathrm{kyr}$ of evolution.
A common feature of these runs is a merger occurring relatively early
(at $\simeq$$40\,\mathrm{kyr}$ and $\simeq$$51\,\mathrm{kyr}$ in simulations I and II, respectively),
followed by a second merger 
(at $\simeq$$81\,\mathrm{kyr}$ and $\simeq$$94\,\mathrm{kyr}$ in simulations I and II, respectively).
The simulation III covers $180\,\mathrm{kyr}$ of evolution.
There is a late violent sequence at $\simeq150\,\mathrm{kyr}$ during which 2 mergers occur
and 3 embryos are scattered out of the simulated part of the disk.

In simulations I--III, only transient binaries are formed.
However, we detect one case of a hardened binary
in the simulation IV which covers $120\,\mathrm{kyr}$ of evolution.
The simulation also contains 190 transients
compared to 65 cases found in our nominal simulation
which implies that the increased multiplicity of the system (8 instead
of 4 embryos) logically increases the frequency of embryo encounters.

Figure~\ref{fig:verif_binaries} shows the evolution of orbital elements
of the binary configurations participating in binary hardening. 
First, a transient consisting of embryos $2$ and $7$ forms at about $122.5\,\mathrm{kyr}$.
After $\simeq1\,\mathrm{kyr}$, it undergoes a 3-body encounter during which
the configuration changes to embryos $2$ and $6$ and the binary semimajor
axis decreases. Then the binary evolves for about $5\,\mathrm{kyr}$
due to pebble accretion and additional 3-body encounters. 
The secular rate of change of $a$ is clearly related to the 
value of $I$, suggesting that the deposition of pebbles onto a prograde
binary causes the separation to decrease faster compared to a retrograde case.
As in the nominal simulation, the binary separation shrinks until the binary merges.

Although the statistics of our 5 (1 nominal and 4 additional) RHD simulations
is still poor, we can nevertheless conclude that a binary planet
with considerable lifetime can form at least in some cases.
We can also make a crude estimate of the formation efficiency
$f_{\mathrm{form}}$ which we define as a fraction of
simulations in which a binary planet was formed and then hardened,
obtaining $f_{\mathrm{form}}\simeq0.4$.

For simulations which formed such binaries,
we also define the total time $\tau_{\mathrm{bp}}$ for
which the binary planet existed in the system
(regardless of which embryos were bound). The binary
in our nominal simulation existed for $\simeq$$16\,\mathrm{kyr}$ and the
binary in the simulation IV existed for $\simeq$$6\,\mathrm{kyr}$ in total.
Taking the arithmetic mean of these two values we obtain
$\tau_{\mathrm{bp}}\simeq11\,\mathrm{kyr}$.

\section{Discussing binary planets}
\label{sec:discussion}

\subsection{Model complexity and limitations}
\label{sec:model_limitations}

The choice of the RHD model was not a priori
motivated by studying formation of binary planets.
The initial motivation was to check how the evolution
described in \cite{Chrenko_etal_2017A&A...606A.114C}
changes if the orbits of embryos become inclined due to
the vertical hot-trail effect
\citep[only the horizontal hot-trail was modelled in][]{Chrenko_etal_2017A&A...606A.114C}.
We found binary planet formation as an unexpected
yet natural outcome of the model.

It is possible that a less complex model could be applied
to scan the parametric space, e.g. by N-body integrations.
But although there are many state-of-the-art
N-body models of migration in multiplanet systems with disks
\citep[e.g.][]{Cossou_etal_2013A&A...553L...2C,Cossou_etal_2014A&A...569A..56C,Izidoro_etal_2017MNRAS.470.1750I},
none of them (to our knowledge) identified formation of binary planets.
This suggests that there might be issues
preventing binary planet formation in such models.

We demonstrated in Section~\ref{sec:transient_formation}
that hydrodynamic effects are important during formation
of transient binaries by 2-body encounters. The pair
of approaching embryos creates perturbations in the disk
which differ from perturbations arising from isolated
embryos. The prescriptions for the
disk torques which are currently used in N-body models
\citep{Paardekooper_etal_2010MNRAS.401.1950P,Paardekooper_etal_2011MNRAS.410..293P,Cresswell_Nelson_2008A&A...482..677C,Fendyke_Nelson_2014MNRAS.437...96F}
cannot account for such effects because they were derived
from models containing a single embryo, not an interacting
pair. Moreover, such prescriptions do not account for the
thermal torque and hot-trail effect.

We point out that our RHD model has some limitations as well.
First, although the grid resolution leads to
  numerical convergence of the migration rate for low-mass planets
  \citep[see e.g.][for resolution discussions]{Lega_etal_2014MNRAS.440..683L,Broz_etal_2018arXiv181003385B},
it is not tuned to resolve binary configurations with the smallest orbital separations well enough.
However, looking at the turquoise curve in Figure~\ref{fig:atetit_ISOL_binary}
(i.e. the case without additional perturbers), the marginal
change of orbital elements indicates very low level of numerical dissipation,
safely negligible over a typical binary lifetime during the disk phase.
Second, the vertical hot-trail cannot be implemented
in our 2D model in a self-consistent way.
3D simulations would be required to assess the importance of the
vertical dimension which we neglect here.
Finally, the magnitude of the thermal torques depends on the
  thermal diffusivity and therefore on the opacity
  \citep{Masset_2017MNRAS.472.4204M}. However, we used a single
  opacity prescription and it remains unclear if the described
  effects work the same way e.g. in a low-density disk undergoing
photoevaporation.

\subsection{Formation mechanism}

We found that 2-body encounters of planetary
embryos in the gas disk can establish transient binaries.
A binary planet with a considerable lifetime
can form from a transient by a 3-body encounter
which provides the necessary energy dissipation to
make the pair more tightly bound.
Here we discuss whether there are other mechanisms
suitable for formation of binary planets.

Additional possibility 
exists during the reaccumulation phase
after a large impact of two embryos approaching
on initially unbound trajectories,
as discussed by \cite{Ryan_etal_2014DPS....4620102R}.
But this situation is highly unlikely. Head-on collisions
usually disrupt the protoplanets
in a way that the reaccumulation forms
a large primary and a low-mass disk,
from which a satellite can be assembled but
not a binary companion.
Only a special grazing geometry with
a large impact parameter can be successful
\citep{Ryan_etal_2014DPS....4620102R}
and it can only produce binaries with
separations of a few planetary radii
due to the angular momentum deficit
of such an encounter.

Finally, planets can be captured in a binary
configuration by means of the tidal dissipation.
\cite{Ochiai_etal_2014ApJ...790...92O} studied
the evolution of three hot Jupiters around a host star
and discovered that binary gas giants can form in $\sim$$10\%$
of systems which undergo orbital crossings.

\subsection{Mass of the binary}

In our hydrodynamic simulations,
the components of binaries have comparable masses
of several $M_{\earth}$
before they merge. But as we found in the follow-up gas-free
N-body simulations, the system often stabilizes
by a collision of one of the embryos 
onto the binary. If the binary survives the collision,
the mass ratio of the components increases to $2/1$.
Therefore it seems that if born from a population of equal mass embryos
(as obtained in the oligarchic growth scenarios), binary
planets would preferentially exist with the component
mass ratios $1/1$ or $2/1$.
This aspect of our model is related to the
choice of the initial embryo masses and is
of course too simplified to capture
the outcome of models where accretion
creates a range of embryo masses.

Although simulations with gas accretion onto the planets
are beyond the scope of our paper, we believe
that runaway accretion of gas onto the binary would disrupt it.
We thus expect the binary planets formed in 3-body
encounters cannot exceed the
masses of giant planet cores. This could be ensured
by the mechanism of pebble isolation
\citep{Lambrechts_etal_2014A&A...572A..35L,Bitsch_etal_2018A&A...612A..30B}
or simply because binaries could form late,
just before the gas disk dispersal.
However, it is possible that binary giant planets
form later by the mechanism of tidal capture \citep{Ochiai_etal_2014ApJ...790...92O}.

\subsection{Tidal evolution}

Orbital evolution due to tidal dissipation is without any doubts
an important factor for the stability
of binary planets. However, it is difficult
to assess the tidal effects at this stage because
there are many unknown parameters.
Large uncertainty lies in the $k_{2}/Q$ parameter,
where $k_{2}$ is the degree 2 Love number and $Q$ is
the tidal quality factor \citep[e.g.][]{Harris_Ward_1982AREPS..10...61H}.
These parameters reflect the interior structure and thus they depend on the
planetary composition (water-rich vs silicate-rich) and state (cold
vs magma worlds), the latter of which
changes on an uncertain timescale.
Moreover, our model treats the planets as point-mass objects,
therefore we have no information about their rotation
which is important to determine the level
of spin-orbital synchronicity.
Additionally, similar masses of the binary components and
their (possibly) retrograde and highly eccentric
orbit make the analysis
of tides even more complicated.
For all these reasons, the model of tidal evolution
of a binary planet should be sufficiently complex
and should account for the internal structure and rheology
\citep[e.g.][]{Boue_etal_2016CeMDA.126...31B,Walterova_Behounkova_2017CeMDA.129..235W}.

\subsection{Occurrence rate and observability}

We define the binary planet occurrence rate $f_{\mathrm{bp}}$
as the fraction of planetary systems which are
expected to contain at least one binary planet hardened by 3-body encounters.
An order-of-magnitude estimate can be obtained
by dividing the time scale $\tau_{\mathrm{bp}}\simeq 10^{4}\,\mathrm{yr}$ for which these binary
planets are typically present in our simulations (Section~\ref{sec:formation_efficiency})
by the lifetime of protoplanetary disks $\tau_{\mathrm{disk}}\simeq 10^{7}\,\mathrm{yr}$
\citep[e.g.][]{Fedele_etal_2010A&A...510A..72F},
correcting for the formation efficiency $f_{\mathrm{form}}\simeq 0.4$ (Section~\ref{sec:formation_efficiency})
and for the fraction of the 
emerging planetary systems $f_{\mathrm{stab}}$
in which the binary planet can survive after the
gas disk dispersal (Section~\ref{sec:nbody_models}). This leads to
\begin{equation}
  f_{\mathrm{bp}} = f_{\mathrm{form}}\frac{\tau_{\mathrm{bp}}}{\tau_{\mathrm{disk}}}f_{\mathrm{stab}}  \, .
  \label{eq:fraction_of_bp}
\end{equation}

We quantify two characteristic values for young planetary systems
(using $f_{\mathrm{stab},10\,\mathrm{Myr}}\simeq0.15$) and for old
planetary systems (using $f_{\mathrm{stab},4.5\,\mathrm{Gyr}}\simeq0.04$), leading to
$f_{\mathrm{bp},10\,\mathrm{Myr}}\simeq6\times10^{-5}$ and $f_{\mathrm{bp},4.5\,\mathrm{Gyr}}\simeq2\times10^{-5}$.
In other words, one out of $\simeq$$2$--$5\times10^{4}$ planetary systems should contain a binary planet
formed by 2- and 3-body encounters.

We emphasize that the estimate is highly uncertain
because it is inferred from a small
number of tests. Moreover, our simulations cover
only a small region of the protoplanetary disk, the evolution
time scale is still short compared to the disk's lifetime,
and the number of embryos is relatively low. Last but not least,
our estimate only assumes that binaries form by the processes
identified in this work.

As pointed out in Section~\ref{sec:intro},  
the estimated occurrence rate
should motivate a systematic search for binary planets
in the observational data.
For example, binary planets should be detectable by the Hunt for Exomoons
with {\em Kepler} (HEK) project \citep{Kipping_etal_2012ApJ...750..115K}.
The sensitivity of this survey is $\simeq$$40\%$
for binaries with Pluto-Charon mass ratios
\citep{Kipping_etal_2015ApJ...813...14K}.

\subsection{Role in planetary systems}
Possible existence of binary planets
opens new avenues in planetary sciences.
A study of long-term orbital dynamics and stability
of binaries in various systems is needed {\citep[e.g.][]{Donnison_2010MNRAS.406.1918D}},
including an assessment of their tidal evolution.
Binary planets are challenging for hydrodynamic modeling
as well. Local high-resolution simulations of
interactions with the disk, pebble accretion, pebble
isolation {\citep[e.g.][]{Bitsch_etal_2018A&A...612A..30B}}
and gas accretion {\citep[e.g.][]{Lambrechts_Lega_2017A&A...606A.146L}}
should be performed for binary planets (preferentially
in 3D) to understand the impact of these processes
in detail.

In this paper, we reported various fates
of binary planets (considering the set
of gas-free N-body simulations).
Frequently, one of the binary components
underwent a collision with an equally large impactor,
or the binary merged.
The merger of the binary planet
is a process that should be investigated
\citep[e.g. by the smoothed-particle hydrodynamics (SPH) method,][]{Jutzi_2015P&SS..107....3J}.

A merger can occur in a situation
when the binary orbit is inclined 
with respect to the global orbital plane
and the resulting body would then retain
the initial angular momentum of the binary,
forming a planet (or a
giant-planet core) with an
angular tilt of the rotational axis. It might
be worth investigating a relation of such an
event to the origin of Uranus.

Moreover, the collision of the binary components
would statistically occur on high impact angles.
It is interesting that such impact
angles and similar masses of the target and the impactor
(although on unbound trajectories) were also used
for successful explanation of the impact
origin of the Earth-Moon system \citep{Canup_2012Sci...338.1052C}.

\section{Conclusions}
\label{sec:conclusions}

By means of 2D radiation hydrodynamic simulations
with 3D planetary orbits,
we described formation of binary planets
in a system of migrating super-Earths.
A key ingredient of the model is the vertical hot-trail
effect \citep{Eklund_Masset_2017MNRAS.469..206E}
which was incorporated by reducing the efficiency
of the inclination damping
prescription \citep{Tanaka_Ward_2004ApJ...602..388T}.
We also accounted for the pebble disk, pebble
accretion and accretion heating, which
naturally produces the horizontal hot-trail
effect, providing eccentricity excitation
\citep{Chrenko_etal_2017A&A...606A.114C}.

When convergent migration drives the planetary embryos together,
the geometry of their
encounters allows for vertical perturbations owing
to the non-zero inclinations.
The orbits become vertically stirred and dynamically
hot, reaching inclinations up to $\simeq$$2^{\circ}$.

Numerous transient binary planets form during the simulations
by gas-assisted 2-body encounters but such transients
quickly dissolve. Binary planets with longer lifetimes
$\sim$$10^{4}\,\mathrm{yr}$ form when a transient undergoes a 3-body
encounter with a third embryo. During this process of binary
hardening, energy is removed from the binary orbit
and the separation of components decreases.
Also, 3-body encounters often reconfigure the binary
when one of the components swaps places with the encountered embryo.
The existence of hardened binaries in our simulations
typically ends with a merger of its components which forms a giant planet core.

The role of the gas disk for binary planet formation is twofold.
In 2-body encounters, the disk can dissipate orbital
energy of the embryos thus aiding the gravitational capture.
The dissipation is provided when the embryos cross a
shared spiral arm.
Regarding 3-body encounters, the disk torques hold
the embryos closely packed and the hot-trail effect maintains
the eccentricities and inclinations excited, increasing
the probability that a transient will encounter another
embryo before it dissolves.

We conducted numerical experiments
to test the stability and evolution of binary planets
in cases when pebble accretion is halted, 
or accretion heating is inefficient, or the disk dissipates.
We found that pebble accretion causes a secular decrease of $a$
of the binary whereas $e$ increases due to the hot-trail effect.

For the binary to survive after the disk dispersal,
it is required that the surrounding embryos are removed
or reconfigured dynamically.
Quite often, a stable configuration is achieved when
 one of the components of the surviving binary undergoes
a merger with another embryo, increasing the binary mass ratio
from approximately $1/1$ to $2/1$.

We roughly estimated the expected
fraction of planetary systems with binary planets to be
$f_{\mathrm{bp}}\simeq2$--$6\times10^{-5}$,
where the upper limit holds for young planetary systems
and the lower limit holds for $4.5\,\mathrm{Gyr}$ old systems.
In other words, a binary planet should be present
in one planetary system out of $\simeq$$2$--$5\times10^{4}$.

One can think of many new applications
which the possible existence of binary planets brings.
First, although binary planets are yet-to-be
discovered, our occurrence rate estimate is encouraging
for future observations.
Second, the hydrodynamic interactions of binary planets
with the disk may be different compared to a single planet
and are worth investigating, preferably in 3D.
Third, collisional models for planetary bodies
which usually focus on unbound trajectories
should also investigate colliding binaries to assess
the possible outcomes.

\acknowledgments

The work of OC has been supported by Charles University (research program no.
UNCE/SCI/023; project GA UK no. 128216; project SVV-260441)
and by The Ministry of Education, Youth and Sports from the Large
Infrastructures for Research, Experimental Development and
Innovations project „IT4Innovations National Supercomputing Center – LM2015070“.
The  work  of  OC  and  MB  has  been  supported  by  the  Grant
Agency of the Czech Republic (grant no. 13-06083S).
The work of DN has been supported by NASA XRP grant.
Access to computing and storage facilities owned
by parties and projects contributing to the National Grid
Infrastructure MetaCentrum, provided under the programme
“Projects of Large  Research, Development, and Innovations
Infrastructures” (CESNET LM2015042), is greatly appreciated.
We thank Robin Canup and an anonymous referee
for helpful comments on the paper.



\bibliographystyle{aasjournal}
\bibliography{references}

\begin{thebibliography}{}
\expandafter\ifx\csname natexlab\endcsname\relax\def\natexlab#1{#1}\fi
\providecommand{\url}[1]{\href{#1}{#1}}
\providecommand{\dodoi}[1]{doi:~\href{http://doi.org/#1}{\nolinkurl{#1}}}
\providecommand{\doeprint}[1]{\href{http://ascl.net/#1}{\nolinkurl{http://ascl.net/#1}}}
\providecommand{\doarXiv}[1]{\href{https://arxiv.org/abs/#1}{\nolinkurl{https://arxiv.org/abs/#1}}}

\bibitem[{{Araujo} {et~al.}(2008){Araujo}, {Winter}, {Prado}, \& {Vieira
  Martins}}]{Araujo_etal_2008MNRAS.391..675A}
{Araujo}, R.~A.~N., {Winter}, O.~C., {Prado}, A.~F.~B.~A., \& {Vieira Martins},
  R. 2008, \mnras, 391, 675, \dodoi{10.1111/j.1365-2966.2008.13833.x}

\bibitem[{{Astakhov} {et~al.}(2005){Astakhov}, {Lee}, \&
  {Farrelly}}]{Astakhov_etal_2005MNRAS.360..401A}
{Astakhov}, S.~A., {Lee}, E.~A., \& {Farrelly}, D. 2005, \mnras, 360, 401,
  \dodoi{10.1111/j.1365-2966.2005.09072.x}

\bibitem[{{Bailli{\'e}} \&
  {Charnoz}(2014)}]{Baillie_Charnoz_2014ApJ...786...35B}
{Bailli{\'e}}, K., \& {Charnoz}, S. 2014, \apj, 786, 35,
  \dodoi{10.1088/0004-637X/786/1/35}

\bibitem[{{Baruteau} \& {Masset}(2008)}]{Baruteau_Masset_2008ApJ...672.1054B}
{Baruteau}, C., \& {Masset}, F. 2008, \apj, 672, 1054, \dodoi{10.1086/523667}

\bibitem[{{Bell} \& {Lin}(1994)}]{Bell_Lin_1994ApJ...427..987B}
{Bell}, K.~R., \& {Lin}, D.~N.~C. 1994, \apj, 427, 987, \dodoi{10.1086/174206}

\bibitem[{{Ben-Jaffel} \&
  {Ballester}(2014)}]{Ben-Jaffel_Ballester_2014ApJ...785L..30B}
{Ben-Jaffel}, L., \& {Ballester}, G.~E. 2014, \apjl, 785, L30,
  \dodoi{10.1088/2041-8205/785/2/L30}

\bibitem[{{Ben{\'{\i}}tez-Llambay} {et~al.}(2015){Ben{\'{\i}}tez-Llambay},
  {Masset}, {Koenigsberger}, \&
  {Szul{\'a}gyi}}]{Benitez-Llambay_etal_2015Natur.520...63B}
{Ben{\'{\i}}tez-Llambay}, P., {Masset}, F., {Koenigsberger}, G., \&
  {Szul{\'a}gyi}, J. 2015, \nat, 520, 63, \dodoi{10.1038/nature14277}

\bibitem[{{Ben{\'{\i}}tez-Llambay} \&
  {Pessah}(2018)}]{Benitez-Llambay_Pessah_2018ApJ...855L..28B}
{Ben{\'{\i}}tez-Llambay}, P., \& {Pessah}, M.~E. 2018, \apjl, 855, L28,
  \dodoi{10.3847/2041-8213/aab2ae}

\bibitem[{{Bennett} {et~al.}(2014){Bennett}, {Batista}, {Bond}, {Bennett},
  {Suzuki}, {Beaulieu}, {Udalski}, {Donatowicz}, {Bozza}, {Abe}, {Botzler},
  {Freeman}, {Fukunaga}, {Fukui}, {Itow}, {Koshimoto}, {Ling}, {Masuda},
  {Matsubara}, {Muraki}, {Namba}, {Ohnishi}, {Rattenbury}, {Saito}, {Sullivan},
  {Sumi}, {Sweatman}, {Tristram}, {Tsurumi}, {Wada}, {Yock}, {MOA
  Collaboration}, {Albrow}, {Bachelet}, {Brillant}, {Caldwell}, {Cassan},
  {Cole}, {Corrales}, {Coutures}, {Dieters}, {Dominis Prester}, {Fouqu{\'e}},
  {Greenhill}, {Horne}, {Koo}, {Kubas}, {Marquette}, {Martin}, {Menzies},
  {Sahu}, {Wambsganss}, {Williams}, {Zub}, {PLANET Collaboration}, {Choi},
  {DePoy}, {Dong}, {Gaudi}, {Gould}, {Han}, {Henderson}, {McGregor}, {Lee},
  {Pogge}, {Shin}, {Yee}, {{$\mu$}FUN Collaboration}, {Szyma{\'n}ski},
  {Skowron}, {Poleski}, {Koz{\l}owski}, {Wyrzykowski}, {Kubiak},
  {Pietrukowicz}, {Pietrzy{\'n}ski}, {Soszy{\'n}ski}, {Ulaczyk}, {OGLE
  Collaboration}, {Tsapras}, {Street}, {Dominik}, {Bramich}, {Browne},
  {Hundertmark}, {Kains}, {Snodgrass}, {Steele}, {RoboNet Collaboration},
  {Dekany}, {Gonzalez}, {Heyrovsk{\'y}}, {Kandori}, {Kerins}, {Lucas},
  {Minniti}, {Nagayama}, {Rejkuba}, {Robin}, \&
  {Saito}}]{Bennett_etal_2014ApJ...785..155B}
{Bennett}, D.~P., {Batista}, V., {Bond}, I.~A., {et~al.} 2014, \apj, 785, 155,
  \dodoi{10.1088/0004-637X/785/2/155}

\bibitem[{{Birnstiel} {et~al.}(2012){Birnstiel}, {Klahr}, \&
  {Ercolano}}]{Birnstiel_etal_2012A&A...539A.148B}
{Birnstiel}, T., {Klahr}, H., \& {Ercolano}, B. 2012, \aap, 539, A148,
  \dodoi{10.1051/0004-6361/201118136}

\bibitem[{{Bitsch} {et~al.}(2013){Bitsch}, {Crida}, {Morbidelli}, {Kley}, \&
  {Dobbs-Dixon}}]{Bitsch_etal_2013A&A...549A.124B}
{Bitsch}, B., {Crida}, A., {Morbidelli}, A., {Kley}, W., \& {Dobbs-Dixon}, I.
  2013, \aap, 549, A124, \dodoi{10.1051/0004-6361/201220159}

\bibitem[{{Bitsch} {et~al.}(2018){Bitsch}, {Morbidelli}, {Johansen}, {Lega},
  {Lambrechts}, \& {Crida}}]{Bitsch_etal_2018A&A...612A..30B}
{Bitsch}, B., {Morbidelli}, A., {Johansen}, A., {et~al.} 2018, \aap, 612, A30,
  \dodoi{10.1051/0004-6361/201731931}

\bibitem[{{Bou{\'e}} {et~al.}(2016){Bou{\'e}}, {Correia}, \&
  {Laskar}}]{Boue_etal_2016CeMDA.126...31B}
{Bou{\'e}}, G., {Correia}, A.~C.~M., \& {Laskar}, J. 2016, Celestial Mechanics
  and Dynamical Astronomy, 126, 31, \dodoi{10.1007/s10569-016-9708-x}

\bibitem[{{Bro{\v z}} {et~al.}(2018){Bro{\v z}}, {Chrenko}, {Nesvorn{\'y}}, \&
  {Lambrechts}}]{Broz_etal_2018arXiv181003385B}
{Bro{\v z}}, M., {Chrenko}, O., {Nesvorn{\'y}}, D., \& {Lambrechts}, M. 2018,
  ArXiv e-prints.
\newblock \doarXiv{1810.03385}

\bibitem[{{Brown} {et~al.}(2006){Brown}, {van Dam}, {Bouchez}, {Le Mignant},
  {Campbell}, {Chin}, {Conrad}, {Hartman}, {Johansson}, {Lafon}, {Rabinowitz},
  {Stomski}, {Summers}, {Trujillo}, \&
  {Wizinowich}}]{Brown_etal_2006ApJ...639L..43B}
{Brown}, M.~E., {van Dam}, M.~A., {Bouchez}, A.~H., {et~al.} 2006, \apjl, 639,
  L43, \dodoi{10.1086/501524}

\bibitem[{{Brozovi{\'c}} {et~al.}(2015){Brozovi{\'c}}, {Showalter}, {Jacobson},
  \& {Buie}}]{Brozovic_etal_2015Icar..246..317B}
{Brozovi{\'c}}, M., {Showalter}, M.~R., {Jacobson}, R.~A., \& {Buie}, M.~W.
  2015, \icarus, 246, 317, \dodoi{10.1016/j.icarus.2014.03.015}

\bibitem[{{Canup}(2011)}]{Canup_2011AJ....141...35C}
{Canup}, R.~M. 2011, \aj, 141, 35, \dodoi{10.1088/0004-6256/141/2/35}

\bibitem[{{Canup}(2012)}]{Canup_2012Sci...338.1052C}
---. 2012, Science, 338, 1052, \dodoi{10.1126/science.1226073}

\bibitem[{{Chiang} \& {Goldreich}(1997)}]{Chiang_Goldreich_1997ApJ...490..368C}
{Chiang}, E.~I., \& {Goldreich}, P. 1997, \apj, 490, 368

\bibitem[{{Chrenko} {et~al.}(2017){Chrenko}, {Bro{\v z}}, \&
  {Lambrechts}}]{Chrenko_etal_2017A&A...606A.114C}
{Chrenko}, O., {Bro{\v z}}, M., \& {Lambrechts}, M. 2017, \aap, 606, A114,
  \dodoi{10.1051/0004-6361/201731033}

\bibitem[{{Christy} \&
  {Harrington}(1978)}]{Christy_Harrington_1978AJ.....83.1005C}
{Christy}, J.~W., \& {Harrington}, R.~S. 1978, \aj, 83, 1005,
  \dodoi{10.1086/112284}

\bibitem[{{Cordeiro} {et~al.}(1999){Cordeiro}, {Martins}, \&
  {Leonel}}]{Cordeiro_etal_1999AJ....117.1634C}
{Cordeiro}, R.~R., {Martins}, R.~V., \& {Leonel}, E.~D. 1999, \aj, 117, 1634,
  \dodoi{10.1086/300764}

\bibitem[{{Cossou} {et~al.}(2014){Cossou}, {Raymond}, {Hersant}, \&
  {Pierens}}]{Cossou_etal_2014A&A...569A..56C}
{Cossou}, C., {Raymond}, S.~N., {Hersant}, F., \& {Pierens}, A. 2014, \aap,
  569, A56, \dodoi{10.1051/0004-6361/201424157}

\bibitem[{{Cossou} {et~al.}(2013){Cossou}, {Raymond}, \&
  {Pierens}}]{Cossou_etal_2013A&A...553L...2C}
{Cossou}, C., {Raymond}, S.~N., \& {Pierens}, A. 2013, \aap, 553, L2,
  \dodoi{10.1051/0004-6361/201220853}

\bibitem[{{Cresswell} {et~al.}(2007){Cresswell}, {Dirksen}, {Kley}, \&
  {Nelson}}]{Cresswell_etal_2007A&A...473..329C}
{Cresswell}, P., {Dirksen}, G., {Kley}, W., \& {Nelson}, R.~P. 2007, \aap, 473,
  329, \dodoi{10.1051/0004-6361:20077666}

\bibitem[{{Cresswell} \& {Nelson}(2008)}]{Cresswell_Nelson_2008A&A...482..677C}
{Cresswell}, P., \& {Nelson}, R.~P. 2008, \aap, 482, 677,
  \dodoi{10.1051/0004-6361:20079178}

\bibitem[{{Crida} {et~al.}(2008){Crida}, {S{\'a}ndor}, \&
  {Kley}}]{Crida_etal_2008A&A...483..325C}
{Crida}, A., {S{\'a}ndor}, Z., \& {Kley}, W. 2008, \aap, 483, 325,
  \dodoi{10.1051/0004-6361:20079291}

\bibitem[{{D'Angelo} {et~al.}(2003){D'Angelo}, {Henning}, \&
  {Kley}}]{DAngelo_etal_2003ApJ...599..548D}
{D'Angelo}, G., {Henning}, T., \& {Kley}, W. 2003, \apj, 599, 548,
  \dodoi{10.1086/379224}

\bibitem[{{de Val-Borro} {et~al.}(2006){de Val-Borro}, {Edgar}, {Artymowicz},
  {Ciecielag}, {Cresswell}, {D'Angelo}, {Delgado-Donate}, {Dirksen}, {Fromang},
  {Gawryszczak}, {Klahr}, {Kley}, {Lyra}, {Masset}, {Mellema}, {Nelson},
  {Paardekooper}, {Peplinski}, {Pierens}, {Plewa}, {Rice}, {Sch{\"a}fer}, \&
  {Speith}}]{deValBorro_etal_2006MNRAS.370..529D}
{de Val-Borro}, M., {Edgar}, R.~G., {Artymowicz}, P., {et~al.} 2006, \mnras,
  370, 529, \dodoi{10.1111/j.1365-2966.2006.10488.x}

\bibitem[{{Donnison}(2010)}]{Donnison_2010MNRAS.406.1918D}
{Donnison}, J.~R. 2010, \mnras, 406, 1918,
  \dodoi{10.1111/j.1365-2966.2010.16796.x}

\bibitem[{{Eklund} \& {Masset}(2017)}]{Eklund_Masset_2017MNRAS.469..206E}
{Eklund}, H., \& {Masset}, F.~S. 2017, \mnras, 469, 206,
  \dodoi{10.1093/mnras/stx856}

\bibitem[{{Erikson} {et~al.}(2012){Erikson}, {Santerne}, {Renner}, {Barge},
  {Aigrain}, {Alapini}, {Almenara}, {Alonso}, {Auvergne}, {Baglin}, {Benz},
  {Bonomo}, {Bord{\'e}}, {Bouchy}, {Bruntt}, {Cabrera}, {Carone}, {Carpano},
  {Csizmadia}, {Deleuil}, {Deeg}, {D{\'{\i}}az}, {Dvorak}, {Ferraz-Mello},
  {Fridlund}, {Gandolfi}, {Gazzano}, {Gillon}, {Guenther}, {Guillot}, {Hatzes},
  {H{\'e}brard}, {Jorda}, {Lammer}, {L{\'e}ger}, {Llebaria}, {Mayor}, {Mazeh},
  {Moutou}, {Ollivier}, {Ofir}, {P{\"a}tzold}, {Pepe}, {Pont}, {Queloz},
  {Rabus}, {Rauer}, {R{\'e}gulo}, {Rouan}, {Samuel}, {Schneider}, {Shporer},
  {Tingley}, {Udry}, \& {Wuchterl}}]{Erikson_etal_2012A&A...539A..14E}
{Erikson}, A., {Santerne}, A., {Renner}, S., {et~al.} 2012, \aap, 539, A14,
  \dodoi{10.1051/0004-6361/201116934}

\bibitem[{{Fedele} {et~al.}(2010){Fedele}, {van den Ancker}, {Henning},
  {Jayawardhana}, \& {Oliveira}}]{Fedele_etal_2010A&A...510A..72F}
{Fedele}, D., {van den Ancker}, M.~E., {Henning}, T., {Jayawardhana}, R., \&
  {Oliveira}, J.~M. 2010, \aap, 510, A72, \dodoi{10.1051/0004-6361/200912810}

\bibitem[{{Fendyke} \& {Nelson}(2014)}]{Fendyke_Nelson_2014MNRAS.437...96F}
{Fendyke}, S.~M., \& {Nelson}, R.~P. 2014, \mnras, 437, 96,
  \dodoi{10.1093/mnras/stt1867}

\bibitem[{{Fraser} {et~al.}(2017){Fraser}, {Bannister}, {Pike}, {Marsset},
  {Schwamb}, {Kavelaars}, {Lacerda}, {Nesvorn{\'y}}, {Volk}, {Delsanti},
  {Benecchi}, {Lehner}, {Noll}, {Gladman}, {Petit}, {Gwyn}, {Chen}, {Wang},
  {Alexandersen}, {Burdullis}, {Sheppard}, \&
  {Trujillo}}]{Fraser_etal_2017NatAs...1E..88F}
{Fraser}, W.~C., {Bannister}, M.~T., {Pike}, R.~E., {et~al.} 2017, Nature
  Astronomy, 1, 0088, \dodoi{10.1038/s41550-017-0088}

\bibitem[{{Giuppone} {et~al.}(2012){Giuppone}, {Ben{\'{\i}}tez-Llambay}, \&
  {Beaug{\'e}}}]{Giuppone_etal_2012MNRAS.421..356G}
{Giuppone}, C.~A., {Ben{\'{\i}}tez-Llambay}, P., \& {Beaug{\'e}}, C. 2012,
  \mnras, 421, 356, \dodoi{10.1111/j.1365-2966.2011.20310.x}

\bibitem[{{Goldreich} {et~al.}(2002){Goldreich}, {Lithwick}, \&
  {Sari}}]{Goldreich_etal_2002Natur.420..643G}
{Goldreich}, P., {Lithwick}, Y., \& {Sari}, R. 2002, \nat, 420, 643,
  \dodoi{10.1038/nature01227}

\bibitem[{{Han}(2008)}]{Han_2008ApJ...684..684H}
{Han}, C. 2008, \apj, 684, 684, \dodoi{10.1086/590331}

\bibitem[{{Harris} \& {Ward}(1982)}]{Harris_Ward_1982AREPS..10...61H}
{Harris}, A.~W., \& {Ward}, W.~R. 1982, Annual Review of Earth and Planetary
  Sciences, 10, 61, \dodoi{10.1146/annurev.ea.10.050182.000425}

\bibitem[{{Heller}(2014)}]{Heller_2014ApJ...787...14H}
{Heller}, R. 2014, \apj, 787, 14, \dodoi{10.1088/0004-637X/787/1/14}

\bibitem[{{Heller}(2018)}]{Heller_2018A&A...610A..39H}
---. 2018, \aap, 610, A39, \dodoi{10.1051/0004-6361/201731760}

\bibitem[{{Hills}(1975)}]{Hills_1975AJ.....80..809H}
{Hills}, J.~G. 1975, \aj, 80, 809, \dodoi{10.1086/111815}

\bibitem[{{Hills}(1990)}]{Hills_1990AJ.....99..979H}
---. 1990, \aj, 99, 979, \dodoi{10.1086/115388}

\bibitem[{{Hippke}(2015)}]{Hippke_2015ApJ...806...51H}
{Hippke}, M. 2015, \apj, 806, 51, \dodoi{10.1088/0004-637X/806/1/51}

\bibitem[{{Hubeny}(1990)}]{Hubeny_1990ApJ...351..632H}
{Hubeny}, I. 1990, \apj, 351, 632, \dodoi{10.1086/168501}

\bibitem[{{Izidoro} {et~al.}(2017){Izidoro}, {Ogihara}, {Raymond},
  {Morbidelli}, {Pierens}, {Bitsch}, {Cossou}, \&
  {Hersant}}]{Izidoro_etal_2017MNRAS.470.1750I}
{Izidoro}, A., {Ogihara}, M., {Raymond}, S.~N., {et~al.} 2017, \mnras, 470,
  1750, \dodoi{10.1093/mnras/stx1232}

\bibitem[{{Jutzi}(2015)}]{Jutzi_2015P&SS..107....3J}
{Jutzi}, M. 2015, \planss, 107, 3, \dodoi{10.1016/j.pss.2014.09.012}

\bibitem[{{Kipping}(2009)}]{Kipping_2009MNRAS.392..181K}
{Kipping}, D.~M. 2009, \mnras, 392, 181,
  \dodoi{10.1111/j.1365-2966.2008.13999.x}

\bibitem[{{Kipping} {et~al.}(2012){Kipping}, {Bakos}, {Buchhave},
  {Nesvorn{\'y}}, \& {Schmitt}}]{Kipping_etal_2012ApJ...750..115K}
{Kipping}, D.~M., {Bakos}, G.~{\'A}., {Buchhave}, L., {Nesvorn{\'y}}, D., \&
  {Schmitt}, A. 2012, \apj, 750, 115, \dodoi{10.1088/0004-637X/750/2/115}

\bibitem[{{Kipping} {et~al.}(2015){Kipping}, {Schmitt}, {Huang}, {Torres},
  {Nesvorn{\'y}}, {Buchhave}, {Hartman}, \&
  {Bakos}}]{Kipping_etal_2015ApJ...813...14K}
{Kipping}, D.~M., {Schmitt}, A.~R., {Huang}, X., {et~al.} 2015, \apj, 813, 14,
  \dodoi{10.1088/0004-637X/813/1/14}

\bibitem[{{Klahr} \& {Kley}(2006)}]{Klahr_Kley_2006A&A...445..747K}
{Klahr}, H., \& {Kley}, W. 2006, \aap, 445, 747,
  \dodoi{10.1051/0004-6361:20053238}

\bibitem[{{Kley}(1989)}]{Kley_1989A&A...208...98K}
{Kley}, W. 1989, \aap, 208, 98

\bibitem[{{Kley} {et~al.}(2009){Kley}, {Bitsch}, \&
  {Klahr}}]{Kley_etal_2009A&A...506..971K}
{Kley}, W., {Bitsch}, B., \& {Klahr}, H. 2009, \aap, 506, 971,
  \dodoi{10.1051/0004-6361/200912072}

\bibitem[{{Kley} \& {Crida}(2008)}]{Kley_Crida_2008}
{Kley}, W., \& {Crida}, A. 2008, \aap, 487, L9,
  \dodoi{10.1051/0004-6361:200810033}

\bibitem[{{Lambrechts} \&
  {Johansen}(2012)}]{Lambrechts_Johansen_2012A&A...544A..32L}
{Lambrechts}, M., \& {Johansen}, A. 2012, \aap, 544, A32,
  \dodoi{10.1051/0004-6361/201219127}

\bibitem[{{Lambrechts} \&
  {Johansen}(2014)}]{Lambrechts_Johansen_2014A&A...572A.107L}
---. 2014, \aap, 572, A107, \dodoi{10.1051/0004-6361/201424343}

\bibitem[{{Lambrechts} {et~al.}(2014){Lambrechts}, {Johansen}, \&
  {Morbidelli}}]{Lambrechts_etal_2014A&A...572A..35L}
{Lambrechts}, M., {Johansen}, A., \& {Morbidelli}, A. 2014, \aap, 572, A35,
  \dodoi{10.1051/0004-6361/201423814}

\bibitem[{{Lambrechts} \& {Lega}(2017)}]{Lambrechts_Lega_2017A&A...606A.146L}
{Lambrechts}, M., \& {Lega}, E. 2017, \aap, 606, A146,
  \dodoi{10.1051/0004-6361/201731014}

\bibitem[{{Laughlin} \&
  {Chambers}(2002)}]{Laughling_Chambers_2002AJ....124..592L}
{Laughlin}, G., \& {Chambers}, J.~E. 2002, \aj, 124, 592,
  \dodoi{10.1086/341173}

\bibitem[{{Lee} \& {Peale}(2006)}]{Lee_Peale_2006Icar..184..573L}
{Lee}, M.~H., \& {Peale}, S.~J. 2006, \icarus, 184, 573,
  \dodoi{10.1016/j.icarus.2006.04.017}

\bibitem[{{Lega} {et~al.}(2014){Lega}, {Crida}, {Bitsch}, \&
  {Morbidelli}}]{Lega_etal_2014MNRAS.440..683L}
{Lega}, E., {Crida}, A., {Bitsch}, B., \& {Morbidelli}, A. 2014, \mnras, 440,
  683, \dodoi{10.1093/mnras/stu304}

\bibitem[{{Lega} {et~al.}(2013){Lega}, {Morbidelli}, \&
  {Nesvorn{\'y}}}]{Lega_etal_2013MNRAS.431.3494L}
{Lega}, E., {Morbidelli}, A., \& {Nesvorn{\'y}}, D. 2013, \mnras, 431, 3494,
  \dodoi{10.1093/mnras/stt431}

\bibitem[{{Lewis} {et~al.}(2015){Lewis}, {Ochiai}, {Nagasawa}, \&
  {Ida}}]{Lewis_etal_2015ApJ...805...27L}
{Lewis}, K.~M., {Ochiai}, H., {Nagasawa}, M., \& {Ida}, S. 2015, \apj, 805, 27,
  \dodoi{10.1088/0004-637X/805/1/27}

\bibitem[{{Liebig} \&
  {Wambsganss}(2010)}]{Liebig_Wambsganss_2010A&A...520A..68L}
{Liebig}, C., \& {Wambsganss}, J. 2010, \aap, 520, A68,
  \dodoi{10.1051/0004-6361/200913844}

\bibitem[{{Marchis} {et~al.}(2008){Marchis}, {Descamps}, {Baek}, {Harris},
  {Kaasalainen}, {Berthier}, {Hestroffer}, \&
  {Vachier}}]{Marchis_etal_2008Icar..196...97M}
{Marchis}, F., {Descamps}, P., {Baek}, M., {et~al.} 2008, \icarus, 196, 97,
  \dodoi{10.1016/j.icarus.2008.03.007}

\bibitem[{{Marchis} {et~al.}(2006){Marchis}, {Hestroffer}, {Descamps},
  {Berthier}, {Bouchez}, {Campbell}, {Chin}, {van Dam}, {Hartman}, {Johansson},
  {Lafon}, {Le Mignant}, {de Pater}, {Stomski}, {Summers}, {Vachier},
  {Wizinovich}, \& {Wong}}]{Marchis_etal_2006Natur.439..565M}
{Marchis}, F., {Hestroffer}, D., {Descamps}, P., {et~al.} 2006, \nat, 439, 565,
  \dodoi{10.1038/nature04350}

\bibitem[{{Margot} {et~al.}(2002){Margot}, {Nolan}, {Benner}, {Ostro},
  {Jurgens}, {Giorgini}, {Slade}, \&
  {Campbell}}]{Margot_etal_2002Sci...296.1445M}
{Margot}, J.~L., {Nolan}, M.~C., {Benner}, L.~A.~M., {et~al.} 2002, Science,
  296, 1445, \dodoi{10.1126/science.1072094}

\bibitem[{{Masset}(2000)}]{Masset_2000A&AS..141..165M}
{Masset}, F. 2000, \aaps, 141, 165, \dodoi{10.1051/aas:2000116}

\bibitem[{{Masset}(2002)}]{Masset_2002A&A...387..605M}
{Masset}, F.~S. 2002, \aap, 387, 605, \dodoi{10.1051/0004-6361:20020240}

\bibitem[{{Masset}(2017)}]{Masset_2017MNRAS.472.4204M}
---. 2017, \mnras, 472, 4204, \dodoi{10.1093/mnras/stx2271}

\bibitem[{{Masset} \& {Velasco
  Romero}(2017)}]{Masset_VelascoRomero_2017MNRAS.465.3175M}
{Masset}, F.~S., \& {Velasco Romero}, D.~A. 2017, \mnras, 465, 3175,
  \dodoi{10.1093/mnras/stw3008}

\bibitem[{{McKinnon} {et~al.}(2017){McKinnon}, {Stern}, {Weaver}, {Nimmo},
  {Bierson}, {Grundy}, {Cook}, {Cruikshank}, {Parker}, {Moore}, {Spencer},
  {Young}, {Olkin}, {Ennico Smith}, {New Horizons Geology}, {Imaging}, \&
  {Composition Theme Teams}}]{McKinnon_etal_2017Icar..287....2M}
{McKinnon}, W.~B., {Stern}, S.~A., {Weaver}, H.~A., {et~al.} 2017, \icarus,
  287, 2, \dodoi{10.1016/j.icarus.2016.11.019}

\bibitem[{{Menou} \& {Goodman}(2004)}]{Menou_Goodman_2004ApJ...606..520M}
{Menou}, K., \& {Goodman}, J. 2004, \apj, 606, 520, \dodoi{10.1086/382947}

\bibitem[{{Mihalas} \& {Weibel
  Mihalas}(1984)}]{Mihalas_WeibelMihalas_1984frh..book.....M}
{Mihalas}, D., \& {Weibel Mihalas}, B. 1984, {Foundations of radiation
  hydrodynamics} (Oxford University Press, New York)

\bibitem[{{Morbidelli} \&
  {Nesvorn\'y}(2012)}]{Morbidelli_Nesvorny_2012A&A...546A..18M}
{Morbidelli}, A., \& {Nesvorn\'y}, D. 2012, \aap, 546, A18,
  \dodoi{10.1051/0004-6361/201219824}

\bibitem[{{M{\"u}ller} {et~al.}(2012){M{\"u}ller}, {Kley}, \&
  {Meru}}]{Muller_etal_2012A&A...541A.123M}
{M{\"u}ller}, T.~W.~A., {Kley}, W., \& {Meru}, F. 2012, \aap, 541, A123,
  \dodoi{10.1051/0004-6361/201118737}

\bibitem[{{Nesvorn{\'y}} {et~al.}(2010){Nesvorn{\'y}}, {Youdin}, \&
  {Richardson}}]{Nesvorny_etal_2010AJ....140..785N}
{Nesvorn{\'y}}, D., {Youdin}, A.~N., \& {Richardson}, D.~C. 2010, \aj, 140,
  785, \dodoi{10.1088/0004-6256/140/3/785}

\bibitem[{{Noll} {et~al.}(2008){Noll}, {Grundy}, {Stephens}, {Levison}, \&
  {Kern}}]{Noll_etal_2008Icar..194..758N}
{Noll}, K.~S., {Grundy}, W.~M., {Stephens}, D.~C., {Levison}, H.~F., \& {Kern},
  S.~D. 2008, \icarus, 194, 758, \dodoi{10.1016/j.icarus.2007.10.022}

\bibitem[{{Ochiai} {et~al.}(2014){Ochiai}, {Nagasawa}, \&
  {Ida}}]{Ochiai_etal_2014ApJ...790...92O}
{Ochiai}, H., {Nagasawa}, M., \& {Ida}, S. 2014, \apj, 790, 92,
  \dodoi{10.1088/0004-637X/790/2/92}

\bibitem[{{Ormel} \& {Klahr}(2010)}]{Ormel_Klahr_2010A&A...520A..43O}
{Ormel}, C.~W., \& {Klahr}, H.~H. 2010, \aap, 520, A43,
  \dodoi{10.1051/0004-6361/201014903}

\bibitem[{{Paardekooper} {et~al.}(2010){Paardekooper}, {Baruteau}, {Crida}, \&
  {Kley}}]{Paardekooper_etal_2010MNRAS.401.1950P}
{Paardekooper}, S.-J., {Baruteau}, C., {Crida}, A., \& {Kley}, W. 2010, \mnras,
  401, 1950, \dodoi{10.1111/j.1365-2966.2009.15782.x}

\bibitem[{{Paardekooper} {et~al.}(2011){Paardekooper}, {Baruteau}, \&
  {Kley}}]{Paardekooper_etal_2011MNRAS.410..293P}
{Paardekooper}, S.-J., {Baruteau}, C., \& {Kley}, W. 2011, \mnras, 410, 293,
  \dodoi{10.1111/j.1365-2966.2010.17442.x}

\bibitem[{{Paardekooper} \&
  {Mellema}(2008)}]{Paardekooper_Mellema_2008A&A...478..245P}
{Paardekooper}, S.-J., \& {Mellema}, G. 2008, \aap, 478, 245,
  \dodoi{10.1051/0004-6361:20078592}

\bibitem[{{Papaloizou} \&
  {Larwood}(2000)}]{Papaloizou_Larwood_2000MNRAS.315..823P}
{Papaloizou}, J.~C.~B., \& {Larwood}, J.~D. 2000, \mnras, 315, 823,
  \dodoi{10.1046/j.1365-8711.2000.03466.x}

\bibitem[{{Pierens}(2015)}]{Pierens_2015MNRAS.454.2003P}
{Pierens}, A. 2015, \mnras, 454, 2003, \dodoi{10.1093/mnras/stv2024}

\bibitem[{{Pierens} {et~al.}(2013){Pierens}, {Cossou}, \&
  {Raymond}}]{Pierens_etal_2013A&A...558A.105P}
{Pierens}, A., {Cossou}, C., \& {Raymond}, S.~N. 2013, \aap, 558, A105,
  \dodoi{10.1051/0004-6361/201322123}

\bibitem[{{Pravec} {et~al.}(2006){Pravec}, {Scheirich}, {Ku{\v s}nir{\'a}k},
  {{\v S}arounov{\'a}}, {Mottola}, {Hahn}, {Brown}, {Esquerdo}, {Kaiser},
  {Krzeminski}, {Pray}, {Warner}, {Harris}, {Nolan}, {Howell}, {Benner},
  {Margot}, {Gal{\'a}d}, {Holliday}, {Hicks}, {Krugly}, {Tholen}, {Whiteley},
  {Marchis}, {DeGraff}, {Grauer}, {Larson}, {Velichko}, {Cooney}, {Stephens},
  {Zhu}, {Kirsch}, {Dyvig}, {Snyder}, {Reddy}, {Moore}, {Gajdo{\v s}},
  {Vil{\'a}gi}, {Masi}, {Higgins}, {Funkhouser}, {Knight}, {Slivan}, {Behrend},
  {Grenon}, {Burki}, {Roy}, {Demeautis}, {Matter}, {Waelchli}, {Revaz},
  {Klotz}, {Rieugn{\'e}}, {Thierry}, {Cotrez}, {Brunetto}, \&
  {Kober}}]{Pravec_etal_2006Icar..181...63P}
{Pravec}, P., {Scheirich}, P., {Ku{\v s}nir{\'a}k}, P., {et~al.} 2006, \icarus,
  181, 63, \dodoi{10.1016/j.icarus.2005.10.014}

\bibitem[{{Pravec} {et~al.}(2012){Pravec}, {Scheirich}, {Vokrouhlick{\'y}},
  {Harris}, {Ku{\v s}nir{\'a}k}, {Hornoch}, {Pray}, {Higgins}, {Gal{\'a}d},
  {Vil{\'a}gi}, {Gajdo{\v s}}, {Korno{\v s}}, {Oey}, {Hus{\'a}rik}, {Cooney},
  {Gross}, {Terrell}, {Durkee}, {Pollock}, {Reichart}, {Ivarsen}, {Haislip},
  {LaCluyze}, {Krugly}, {Gaftonyuk}, {Stephens}, {Dyvig}, {Reddy}, {Chiorny},
  {Vaduvescu}, {Longa-Pe{\~n}a}, {Tudorica}, {Warner}, {Masi}, {Brinsfield},
  {Gon{\c c}alves}, {Brown}, {Krzeminski}, {Gerashchenko}, {Shevchenko},
  {Molotov}, \& {Marchis}}]{Pravec_etal_2012Icar..218..125P}
{Pravec}, P., {Scheirich}, P., {Vokrouhlick{\'y}}, D., {et~al.} 2012, \icarus,
  218, 125, \dodoi{10.1016/j.icarus.2011.11.026}

\bibitem[{{Rein} \& {Liu}(2012)}]{Rein_Liu_2012A&A...537A.128R}
{Rein}, H., \& {Liu}, S.-F. 2012, \aap, 537, A128,
  \dodoi{10.1051/0004-6361/201118085}

\bibitem[{{Rein} \& {Spiegel}(2015)}]{Rein_Spiegel_2015MNRAS.446.1424R}
{Rein}, H., \& {Spiegel}, D.~S. 2015, \mnras, 446, 1424,
  \dodoi{10.1093/mnras/stu2164}

\bibitem[{{Richardson} \& {Walsh}(2006)}]{Richardson_Walsh_2006AREPS..34...47R}
{Richardson}, D.~C., \& {Walsh}, K.~J. 2006, Annual Review of Earth and
  Planetary Sciences, 34, 47, \dodoi{10.1146/annurev.earth.32.101802.120208}

\bibitem[{{Ryan} {et~al.}(2014){Ryan}, {Nakajima}, \&
  {Stevenson}}]{Ryan_etal_2014DPS....4620102R}
{Ryan}, K., {Nakajima}, M., \& {Stevenson}, D.~J. 2014, in AAS/Division for
  Planetary Sciences Meeting Abstracts, Vol.~46, AAS/Division for Planetary
  Sciences Meeting Abstracts \#46, 201.02

\bibitem[{{Scheeres} {et~al.}(2006){Scheeres}, {Fahnestock}, {Ostro}, {Margot},
  {Benner}, {Broschart}, {Bellerose}, {Giorgini}, {Nolan}, {Magri}, {Pravec},
  {Scheirich}, {Rose}, {Jurgens}, {De Jong}, \&
  {Suzuki}}]{Scheeres_etal_2006Sci...314.1280S}
{Scheeres}, D.~J., {Fahnestock}, E.~G., {Ostro}, S.~J., {et~al.} 2006, Science,
  314, 1280, \dodoi{10.1126/science.1133599}

\bibitem[{{Sim{\'o}} \& {Stuchi}(2000)}]{Simo_Stuchi_2000PhyD..140....1S}
{Sim{\'o}}, C., \& {Stuchi}, T.~J. 2000, Physica D Nonlinear Phenomena, 140, 1,
  \dodoi{10.1016/S0167-2789(99)00211-0}

\bibitem[{{Simon} {et~al.}(2007){Simon}, {Szatm{\'a}ry}, \&
  {Szab{\'o}}}]{Simon_etal_2007A&A...470..727S}
{Simon}, A., {Szatm{\'a}ry}, K., \& {Szab{\'o}}, G.~M. 2007, \aap, 470, 727,
  \dodoi{10.1051/0004-6361:20066560}

\bibitem[{{Simon} {et~al.}(2012){Simon}, {Szab{\'o}}, {Kiss}, \&
  {Szatm{\'a}ry}}]{Simon_etal_2012MNRAS.419..164S}
{Simon}, A.~E., {Szab{\'o}}, G.~M., {Kiss}, L.~L., \& {Szatm{\'a}ry}, K. 2012,
  \mnras, 419, 164, \dodoi{10.1111/j.1365-2966.2011.19682.x}

\bibitem[{{Sonnett} {et~al.}(2015){Sonnett}, {Mainzer}, {Grav}, {Masiero}, \&
  {Bauer}}]{Sonnett_etal_2015ApJ...799..191S}
{Sonnett}, S., {Mainzer}, A., {Grav}, T., {Masiero}, J., \& {Bauer}, J. 2015,
  \apj, 799, 191, \dodoi{10.1088/0004-637X/799/2/191}

\bibitem[{{Tanaka} \& {Ward}(2004)}]{Tanaka_Ward_2004ApJ...602..388T}
{Tanaka}, H., \& {Ward}, W.~R. 2004, \apj, 602, 388, \dodoi{10.1086/380992}

\bibitem[{{Teachey} {et~al.}(2018){Teachey}, {Kipping}, \&
  {Schmitt}}]{Teachey_2018AJ....155...36T}
{Teachey}, A., {Kipping}, D.~M., \& {Schmitt}, A.~R. 2018, \aj, 155, 36,
  \dodoi{10.3847/1538-3881/aa93f2}

\bibitem[{{Veillet} {et~al.}(2002){Veillet}, {Parker}, {Griffin}, {Marsden},
  {Doressoundiram}, {Buie}, {Tholen}, {Connelley}, \&
  {Holman}}]{Veillet_etal_2002Natur.416..711V}
{Veillet}, C., {Parker}, J.~W., {Griffin}, I., {et~al.} 2002, \nat, 416, 711,
  \dodoi{10.1038/416711a}

\bibitem[{{Walker}(1980)}]{Walker_1980MNRAS.192P..47W}
{Walker}, A.~R. 1980, \mnras, 192, 47P, \dodoi{10.1093/mnras/192.1.47P}

\bibitem[{{Walsh} {et~al.}(2008){Walsh}, {Richardson}, \&
  {Michel}}]{Walsh_etal_2008Natur.454..188W}
{Walsh}, K.~J., {Richardson}, D.~C., \& {Michel}, P. 2008, \nat, 454, 188,
  \dodoi{10.1038/nature07078}

\bibitem[{{Walterov{\'a}} \& {B{\v
  e}hounkov{\'a}}(2017)}]{Walterova_Behounkova_2017CeMDA.129..235W}
{Walterov{\'a}}, M., \& {B{\v e}hounkov{\'a}}, M. 2017, Celestial Mechanics and
  Dynamical Astronomy, 129, 235, \dodoi{10.1007/s10569-017-9772-x}

\bibitem[{{Youdin} \& {Lithwick}(2007)}]{Youdin_Lithwick_2007Icar..192..588Y}
{Youdin}, A.~N., \& {Lithwick}, Y. 2007, \icarus, 192, 588,
  \dodoi{10.1016/j.icarus.2007.07.012}

\end{thebibliography}


\listofchanges

\end{document}